\documentclass[10pt]{article}

\usepackage{amsmath}
\usepackage{amssymb}

\usepackage{graphicx}

\usepackage{cite}

\usepackage{color} 

\usepackage[labelfont=bf,labelsep=period,justification=raggedright]{caption}

\makeatletter
\renewcommand{\@biblabel}[1]{\quad#1.}
\makeatother

\date{}

\begin{document}

\begin{flushleft}
{\Large
\textbf{Fluctuation analysis: can estimates be trusted?}
}
\\
Bernard Ycart$^{1,\ast}$
\\
\bf{1} Bernard Ycart 
Laboratoire Jean Kuntzmann, 
Univ. Grenoble-Alpes and CNRS UMR 5224, 
51 rue des Math\'{e}matiques 38041 Grenoble cedex 9, France
\\
$\ast$ E-mail: Bernard.Ycart@imag.fr
\end{flushleft}

\section*{Abstract}
The estimation of mutation probabilities and relative fitnesses
in fluctuation analysis is based on the unrealistic 
hypothesis that the single-cell
times to division are exponentially distributed. 
Using the classical Luria-Delbr\"{u}ck distribution outside its
modelling hypotheses induces an important bias on the
estimation of the relative fitness.
The model is extended here to any division time distribution.
Mutant counts follow
a generalization of the Luria-Delbr\"{u}ck distribution, which depends
on the mean number of mutations, the relative fitness of normal cells
compared to mutants, and the division time distribution of mutant
cells.  Empirical probability generating
function techniques yield precise estimates both of the
mean number of mutations and the
relative fitness of normal cells compared to mutants. In the case
where no information is available on the division time distribution,
it is shown that the estimation procedure using
constant division times yields more reliable results. 
Numerical results both on observed and simulated data are reported.

\section*{Author Summary}
The classical Luria-Delbr\"{u}ck method for estimating 
mutation probabilities and relative fitnesses by fluctuation 
analysis relies upon modelling hypotheses on the time to
division of individual cells which do not fit experimental
observations. Monte-Carlo simulation evidence proves that sizeable
biases can be induced. A new method is proposed that uses more
realistic models and produces more precise estimates. 
\section*{Introduction}
The estimation of mutation parameters in cell growth experiments, or
\emph{fluctuation analysis}, has been
the object of many studies since its introduction by Luria and
Delbr\"{u}ck in 1943 \cite{LuriaDelbruck43}: see reviews by 
Stewart et al. \cite{Stewartetal90}, Angerer
\cite{Angerer01b}, and Foster \cite{Foster06}.
Fluctuation analysis is based on the \emph{Luria-Delbr\"{u}ck} 
distribution, derived under different assumptions by Lea and Coulson 
\cite{LeaCoulson49} and Bartlett
(in the discussion following Armitage
\cite[p.~37]{Armitage52}). 
Mandelbrot \cite{Mandelbrot74}, then Bartlett 
\cite[section 4.31]{Bartlett78} later generalized the Luria-Delbr\"{u}ck
distribution to the differential growth case. 
Since then, fluctuation analysis
with differential growth rates has been advocated by several authors
\cite{Koch82,Jones94,JaegerSarkar95,Zheng02,Zheng05}.
As shown in \cite{HamonYcart12}, Luria-Delbr\"{u}ck distributions 
are made of three ingredients:
\begin{enumerate}
\item 
\emph{The mean number of mutations $\alpha$}, which is the parameter 
of main interest. It is the product of the individual probability of 
mutation by the final number of cells. As already remarked by
Luria and Delbr\"{u}ck \cite[p.~499]{LuriaDelbruck43}, the law of 
small numbers implies that the random number of mutations that occur
during the experiment follows a Poisson distribution with expectation 
$\alpha$.
\item
\emph{The relative fitness $\rho$} 
of normal cells to mutants, i.e. the ratio of the
exponential growth rate of normal cells to that of
mutants. (Growth rate refers here to the constant speed at which
the logarithm of a population of cells grows, not to the size increments of
individual cells). The time
scale does not influence final counts of mutant cells: it may be 
chosen so that the growth rate of mutants
is $1$, in which case $\rho$ is the exponential growth rate of normal
cells. Exponential growth implies that most random mutations occur
rather close to the end of the experiment, and that the time during
which a new mutant clone develops has negative exponential distribution with
parameter $\rho$.
\item
\emph{The random number of cells $M(t)$ in a mutant clone} that develops
for a finite time $t$: it depends crucially on the division times
of mutants. In the classical Luria-Delbr\"{u}ck model, mutants are supposed
to have exponentially distributed division times, which implies that 
$M(t)$ follows the geometric distribution with parameter $\mathrm{e}^{-t}$
(choosing the time scale so that mutants have unit growth rate). 
\end{enumerate}
The first two points can be considered as established
facts: they are in accordance
with experimental data, and grounded on well known probabilistic
results. On the opposite, the hypothesis of exponentially distributed
division times is a purely mathematical convenience and
does not match experimental observations: as remarked as 
early as 1932 by Kelly and Rahn \cite{KellyRahn32,Rahn32}, 
actual division times data are unimodal and right-skewed rather 
than exponential: see \cite{Murphyetal84}. The question investigated
here is: which bias on the estimation of the parameters does the
exponential distribution hypothesis induce, and how can it be reduced?

The ``mathematical convenience'' can be challenged. Admittedly, 
the exponential distribution of division times is the first one
under which a closed mathematical expression for the distribution
of mutants was obtained. Notwithstanding, 
it will be shown that a joint estimation
procedure for $\alpha$ and $\rho$ can be implemented whatever the
distribution of division times. Moreover if the division times
of mutants are supposed to be constant, estimation procedures
are exactly as computationally effective as under the exponential hypothesis.
Since the pioneering observations 
of Kelly and Rahn \cite{KellyRahn32}
progress in experimental settings, from microscopic observation
of single-cell behavior to flow chambers and automated growth
analyzers, has fueled many studies on division times and their 
distributions. Division time data
have been fitted by several types of 
distributions: from Gamma and Log-beta \cite{Kendall48}, to Log-normal
and reciprocal normal \cite{Kubitschek71}: see 
John \cite{John81} and references therein. More recent references
include \cite{Deenicketal03,Stewartetal05,Nivenetal08,Zilmanetal10}. 
There is no such object as ``the'' distribution of division times;
firstly because it would depend not only on the species, strain,
experimental conditions, etc., secondly because many different families of 
distributions can usually fit any given set of observed data. 
I have chosen three families (Gamma, Log-normal,
Inverse Gaussian) and one data set: the historical observations of
Kelly and Rahn on Bacterium aerogenes \cite[Table~2, p.~149]{KellyRahn32}.
A maximum likelihood estimation of the parameters
on the data led to one particular distribution in each family, 
that was rescaled to unit growth rate. 
The three distributions so obtained were considered as realistic 
and used as benchmarks for extensive Monte-Carlo studies.
Samples of size $100$ of generalized Luria-Delbr\"{u}ck
distributions were repeatedly simulated for different values
of $\alpha$ and $\rho$, and for each of the three realistic
distributions. The main conclusion was that using the classical 
Luria-Delbr\"{u}ck distribution estimation procedure yields 
satisfactory results for the estimation of the mean number of mutations
$\alpha$ but introduces a sizeable bias on the estimation of the
relative fitness $\rho$. The estimation procedure that uses
constant division times has a negligible bias and a much better
precision on $\rho$.

I have developed in R \cite{R} a set of functions that output
samples of  generalized Luria-Delbr\"{u}ck 
distributions, compute estimates, confidence regions and p-values for hypothesis
testing. These functions have been made available 
on line\footnote{http://www.ljk.imag.fr/membres/Bernard.Ycart/LD/}.

\section*{Results}
\subsection*{Simulation experiments}
I denote hereafter by $GLD(\alpha,\rho,F)$ the generalized
Luria-Delbr\"uck distribution with parameters $\alpha$, $\rho$, and
$F$: it is the distribution of the final number of mutants in a
fluctuation analysis experiment, when the mean number of mutations is
$\alpha$, the relative fitness of normal cells compared to mutants is
$\rho$, and the distribution of mutant division times is $F$. The
particular case where division times are exponentially distributed
is the classical Luria-Delbr\"{u}ck distribution
$LD(\alpha,\rho)$. Detailed definitions will be given in the
`models and methods' section. 
In real fluctuation analysis experiments, the actual distribution $F$ of
division times is unknown. Therefore the question to be answered was
the following: if a
sample of the  generalized Luria-Delbr\"{u}ck distribution
$GLD(\alpha,\rho,F)$ has been produced, and estimates $\hat{\alpha}$ and
$\hat{\rho}$ are computed from another division time model than $F$,
by how much are these estimates biased, how reliable confidence
intervals on $\alpha$ and $\rho$ can be?

\begin{figure}[!ht]
\begin{center}
\includegraphics[width=4in]{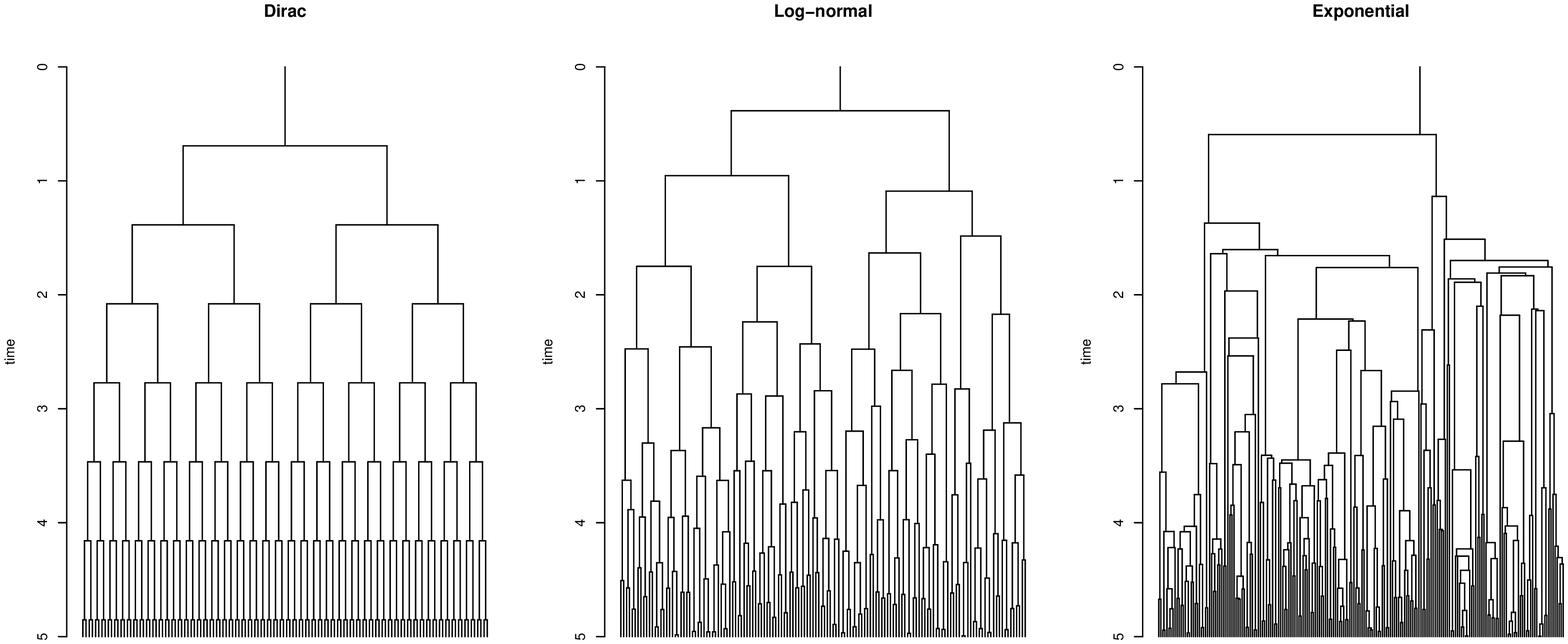}
\end{center}

\caption{
{\bf Clones under of Dirac, Log-normal, and Exponential models.}
The Log-normal distribution has been adjusted on Kelly and Rahn's data.  
All three distribution have been scaled to have unit growth
rates. Clones were simulated up to time 5.
}
\label{fig:clones}
\end{figure}

Three distributions were used in simulation procedures:
Gamma, Log-normal, and Inverse Gaussian; they were adjusted on Kelly
and Rahn's Bacterium aerogenes data. The exact 
definition of the three distributions is detailed 
in the `models and methods' section.
Two models were considered for estimation: the exponential model
(division times follow the negative 
exponential distribution, i.e. the classical model), and the Dirac
model (all division times are equal to the same value). The
corresponding distribution functions are denoted by $F_{\mathrm{exp}}$
and $F_{\mathrm{dir}}$. The estimation procedure is explained in the
`models and methods' section. Figure \ref{fig:clones} represents the
evolution of three
typical clones, simulated with the Dirac model, the Log-normal model,
and the exponential model: the exponential model is much more
irregular than observed in practice: see e.g. \cite[Figure
5]{Beanetal06}.

The simulation study consisted in simulating
samples of the $GLD(\alpha,\rho,F)$, $F$ being a Gamma, Log-normal, or
Inverse Gaussian distribution, then estimating $\alpha$ and $\rho$ as
if $F$ had been $F_{\mathrm{exp}}$ or $F_{\mathrm{dir}}$.
A simulation function for the $GLD(\alpha,\rho,F)$
has been included in the R script made available on line. It was used
to output $10000$ samples of size $100$ for $27$ different sets of
parameters: $\alpha=1,4,8$, $\rho=0.8,1.0,1.2$, $F$ being one of the
three distributions mentioned above.  
Apart from the extensive
study of \cite{Boeetal94}, usual fluctuation experiment 
samples have size of order a few tens, which motivated my choice 
for the sample size. The range of values for $\rho$ is typical of 
practical situations. For $\alpha$, very small values were not considered
as significant: if $\alpha<1$, a large part of the information is
contained in the frequency of zeros: the so called $p_0$-method gives 
almost as good results on $\alpha$ as any other estimator, independently
from the model \cite{HamonYcart12}.

\begin{figure}[!ht]
\begin{center}
\begin{tabular}{ccc}
\includegraphics[width=0.2\linewidth]{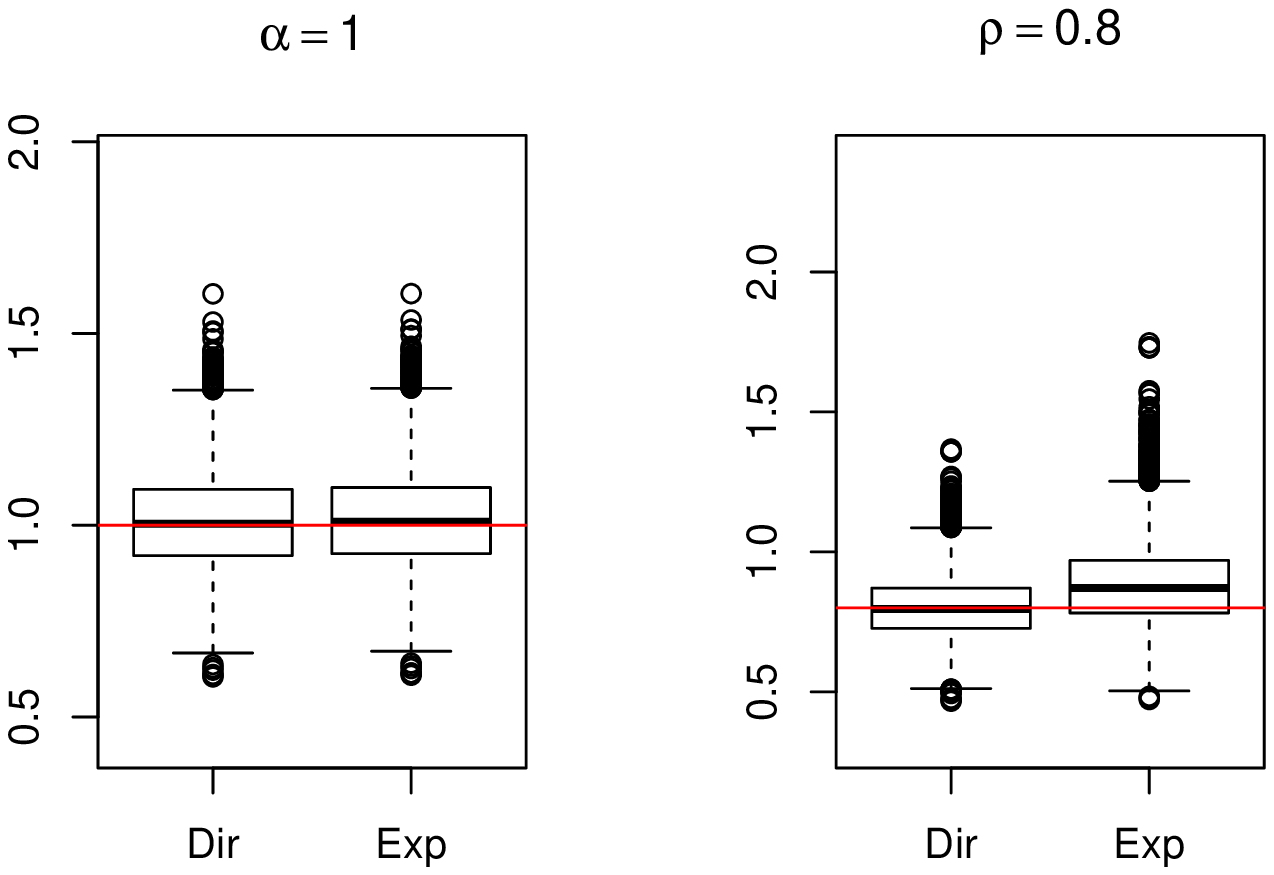}&
\includegraphics[width=0.2\linewidth]{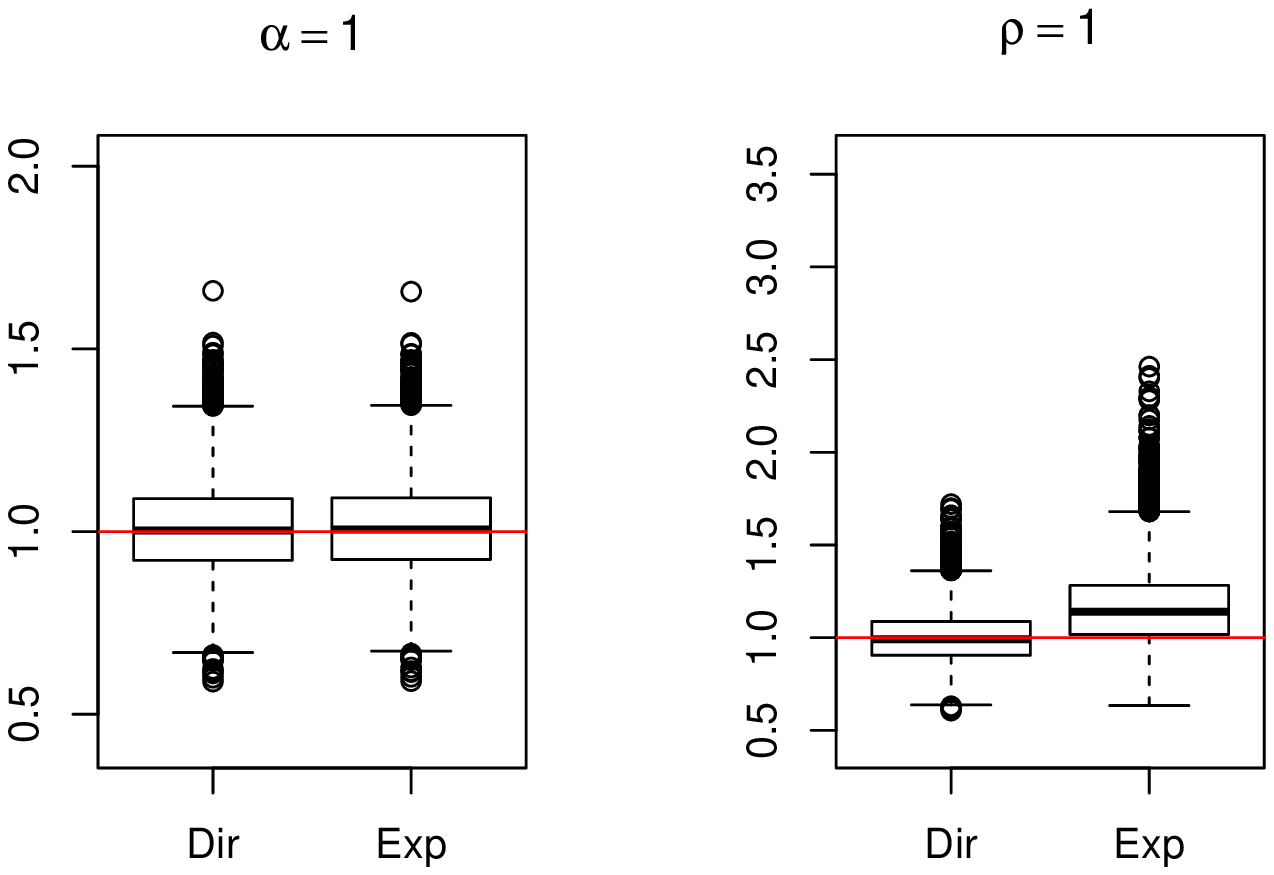}&
\includegraphics[width=0.2\linewidth]{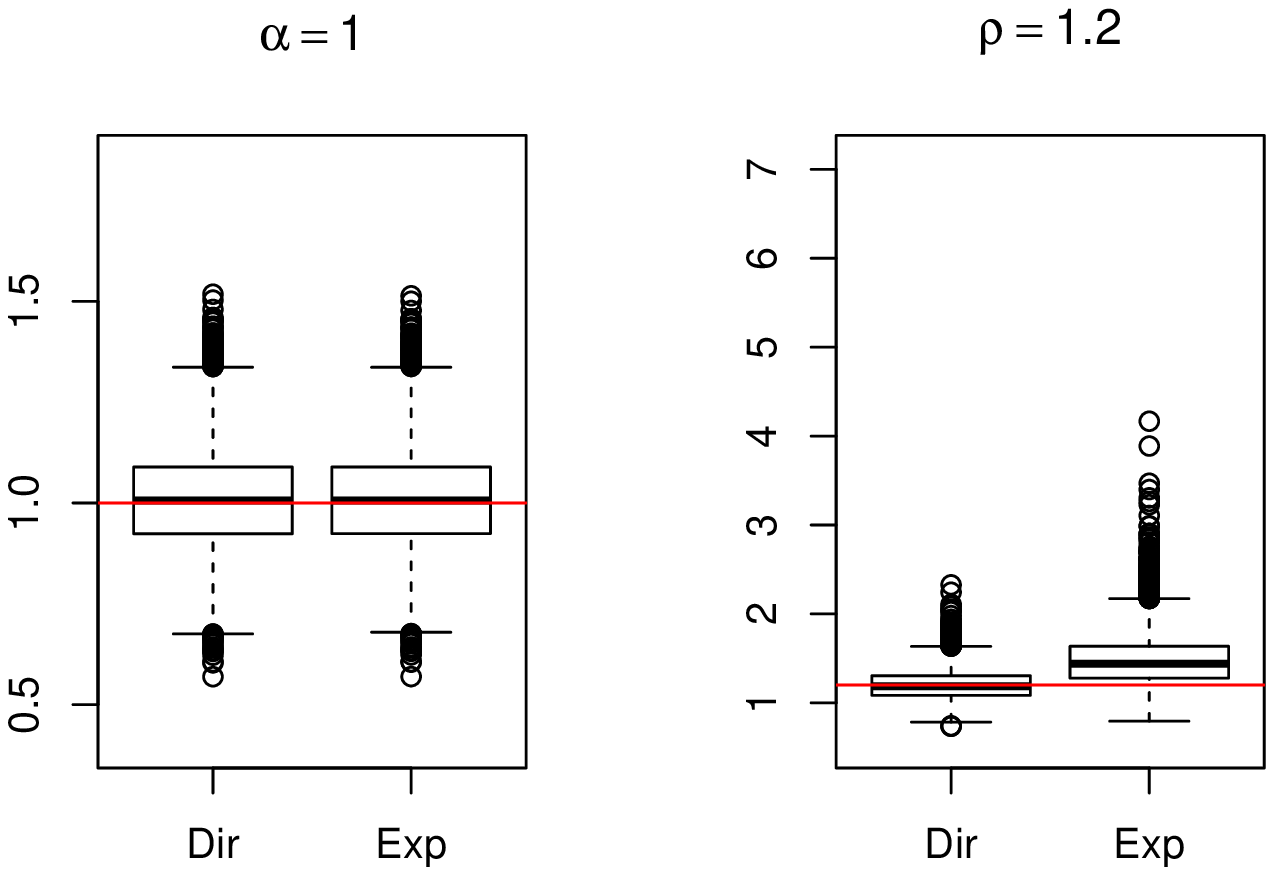}\\
\includegraphics[width=0.2\linewidth]{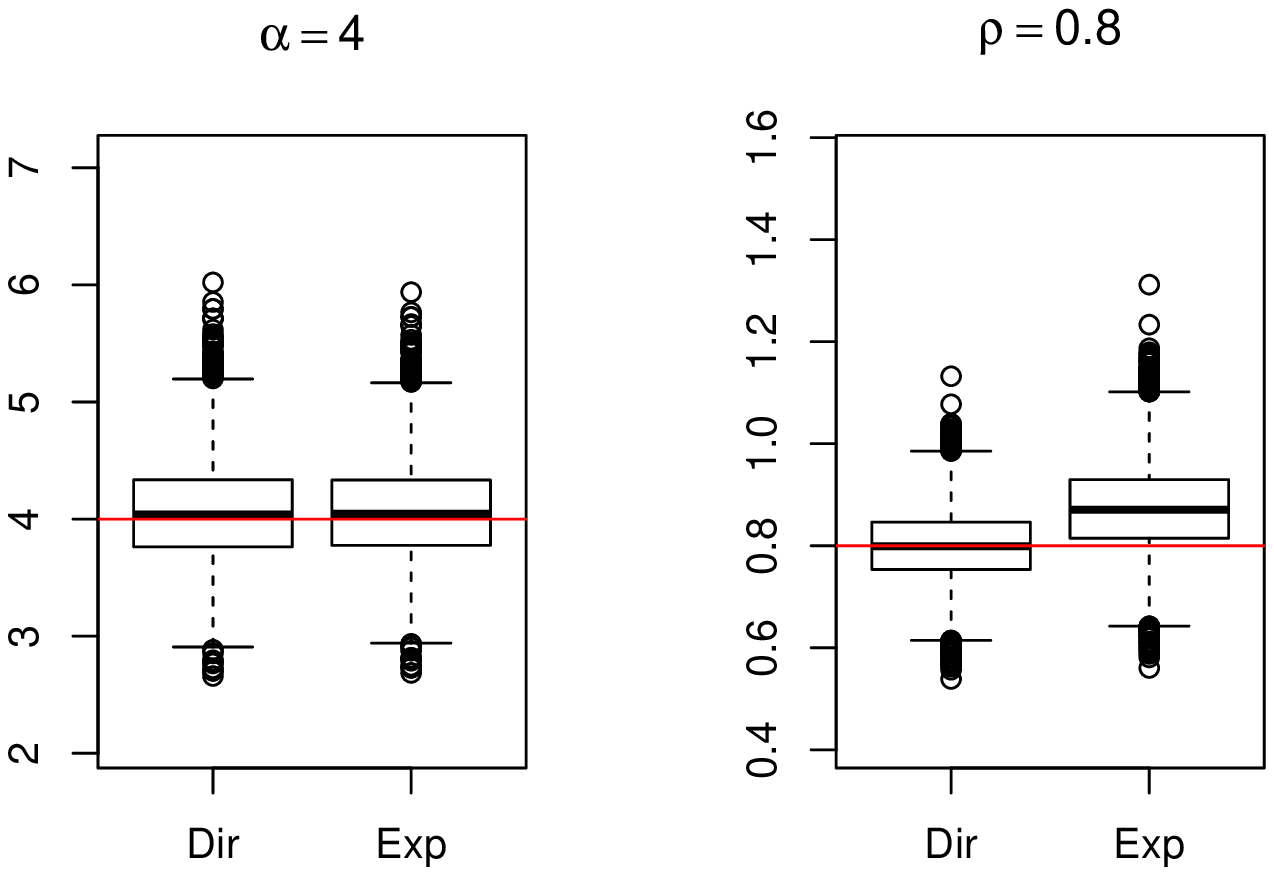}&
\includegraphics[width=0.2\linewidth]{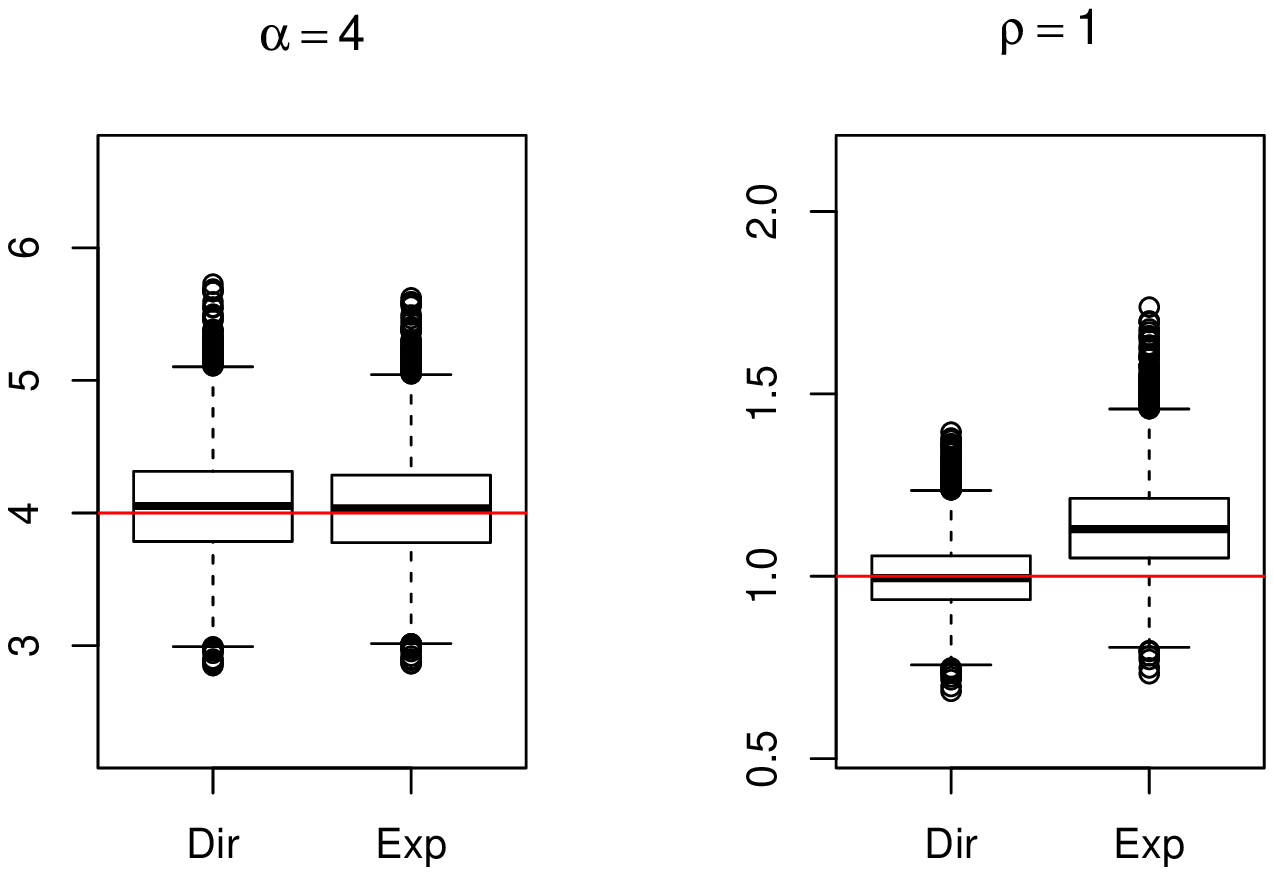}&
\includegraphics[width=0.2\linewidth]{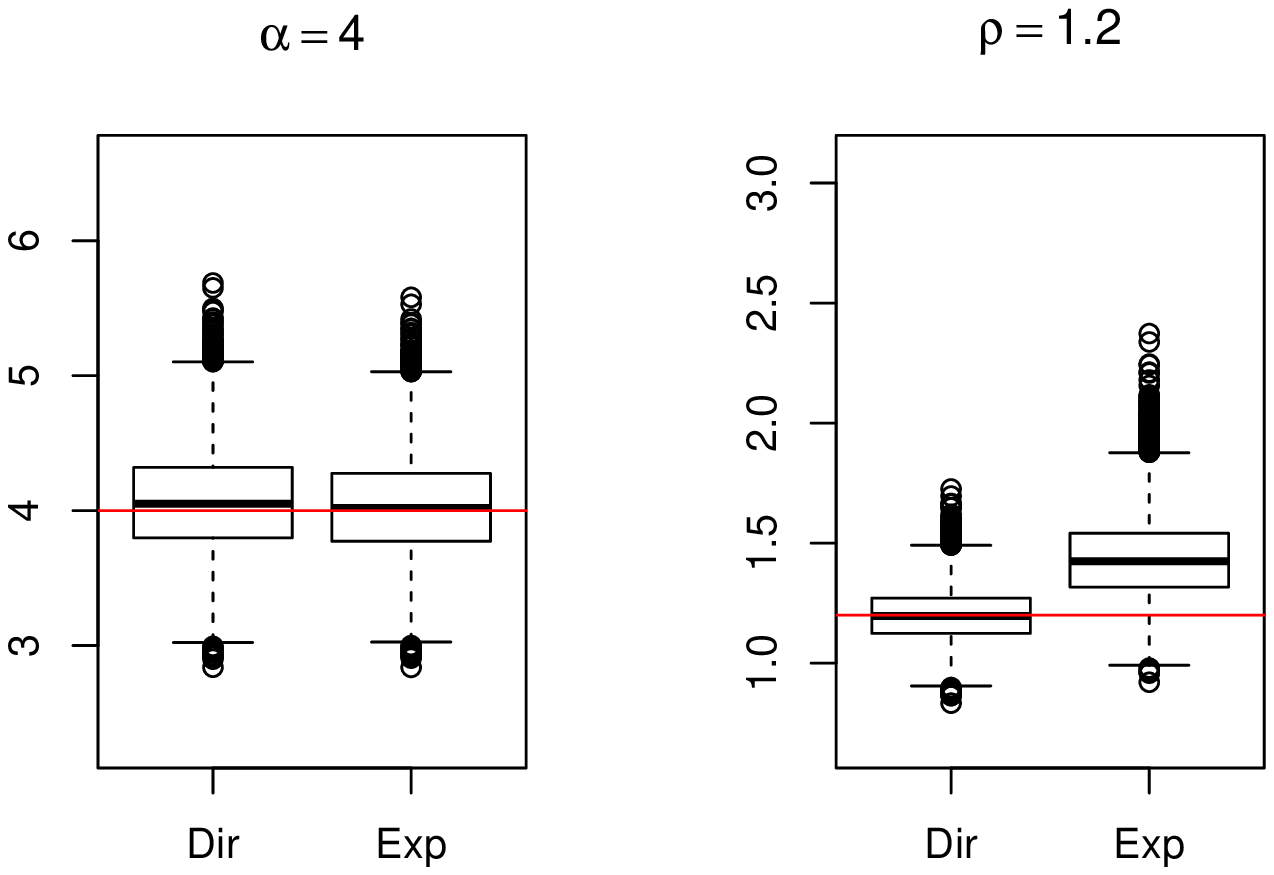}\\
\includegraphics[width=0.2\linewidth]{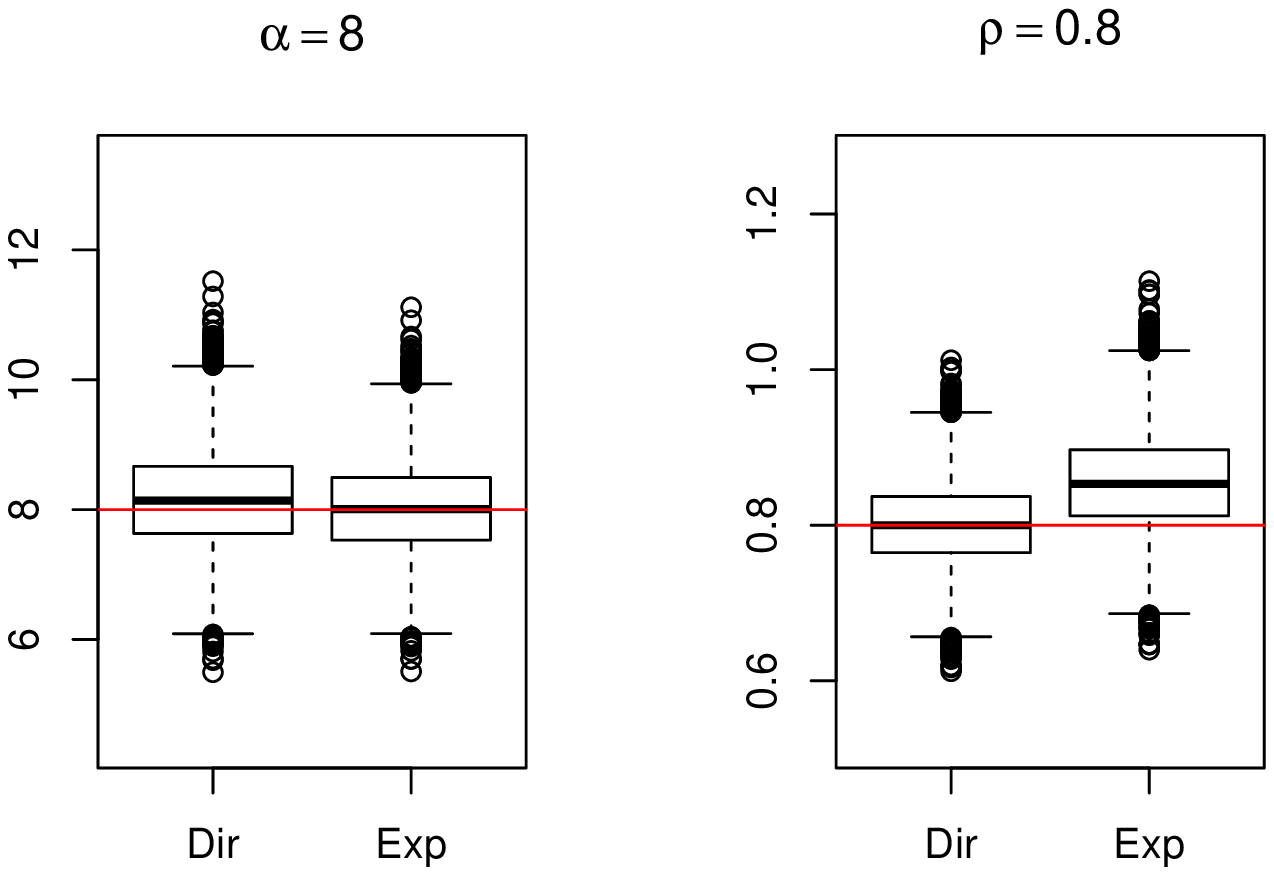}&
\includegraphics[width=0.2\linewidth]{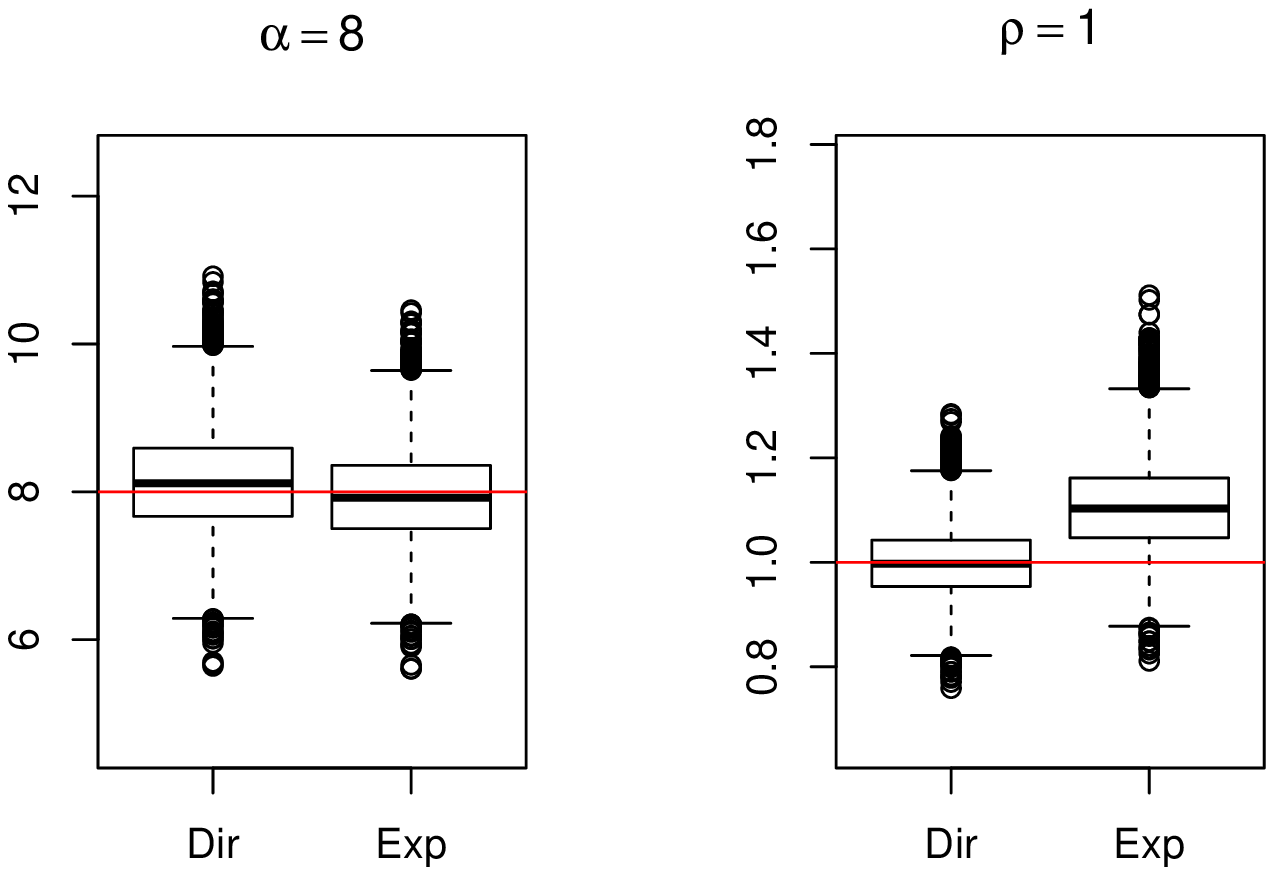}&
\includegraphics[width=0.2\linewidth]{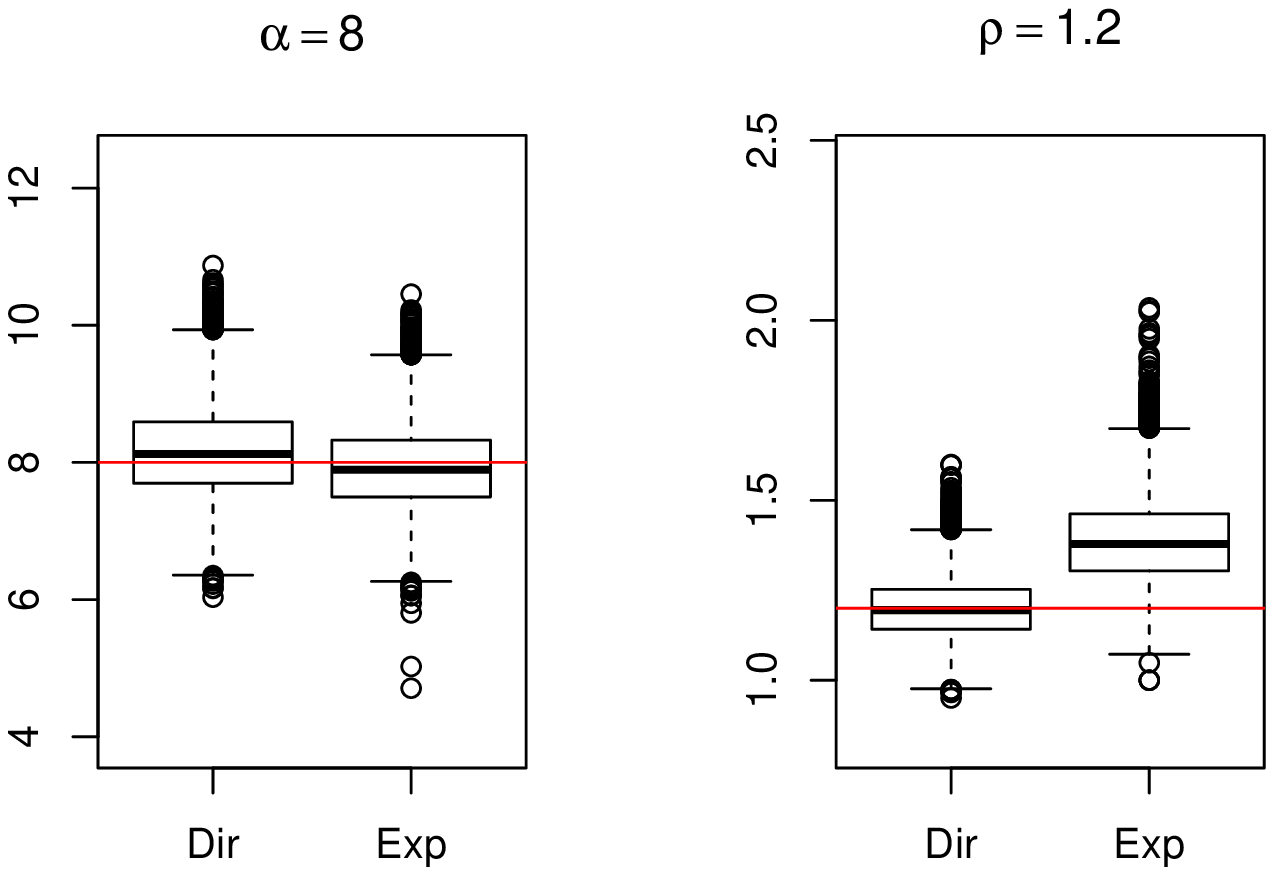}\\
\end{tabular}
\end{center}

\caption{
{\bf Boxplots of estimates of $\alpha$ and $\rho$, using the
  exponential and the Dirac models.} Red horizontal lines mark true
values of the parameters.
For each of the 9 sets of parameters $\alpha=1,4,8$ (rows) and
$\rho=0.8,1.0,1.2$ (columns), 10000 samples of
size 100 of the $GLD(\alpha,\rho,F)$ were simulated, $F$ being the
Log-normal distribution adjusted on Kelly and Rahn's data. The
estimates of $\alpha$ and $\rho$ were calculated with the two models
Dirac and exponential. Each boxplot represents the distribution of the
$10000$ estimates obtained by the Dirac model (left) and the
exponential model (right).
}
\label{fig:boxplots}
\end{figure}

For each of the $270000$
samples, and for the two models $F_{\mathrm{exp}}$ and $F_{\mathrm{dir}}$, 
the estimates of $\alpha$ and $\rho$ were computed, together with
their confidence intervals at level 95\%. The results obtained with 
the three distributions Gamma, Log-normal, and Inverse Gaussian, turned out to
be very similar. Only the results for Log-normal
division times are reported here. Figure \ref{fig:boxplots} displays
the boxplots of the estimated values of $\alpha$ and $\rho$ for
the 9 couples of parameters $\alpha=1,4,8$, $\rho=0.8,1.0,1.2$. 
The following visual observations can be made:
\begin{itemize}
\item the classical exponential model clearly overestimates $\rho$ and
  has a rather large dispersion of estimated values,
\item the Dirac model correctly estimates $\rho$. It induces a 
much smaller bias and dispersion,
\item both models correctly estimate $\alpha$.
\end{itemize}
Further precisions are given in Table \ref{tab:bias}, were mean biases
on 10000 samples are given for each of the 9 couples of
parameters $(\alpha,\rho)$ and the two models.
The mean bias for estimates of $\rho$ using the exponential model
(last column of Table \ref{tab:bias}) is
quite sizeable: between 10\% and 30\% of the true value.

The quality of confidence intervals when the model is not
adapted is illustrated on Table \ref{tab:success}. For each of the
$27000$ samples of size $100$, confidence intervals for $\alpha$ and
$\rho$ at confidence level 95\% 
have been computed using the exponential and Dirac model. Out
of them, a theoretical proportion of
$0.95$ should contain the true value of the estimated parameter. The
proportion of the $10000$ intervals containing the true value has been
computed for each value of the parameters. Table \ref{tab:success}
shows the results for the Log-normal samples (results for the other
two distributions are similar). The confidence intervals for $\alpha$
had a correct proportion of success for both models, slightly better
for estimates using the exponential model. Confidence intervals on
$\rho$ using the Dirac model are also correct. However, the estimation
of $\rho$ using the exponential model was not reliable: up to 30\% of
the 95\% confidence intervals did not contain the true value of
$\rho$ (last column of Table \ref{tab:success}). This result is in
accordance with the strong bias discussed above.

{The parameter of main interest being $\alpha$, the
  results of Tables \ref{tab:bias} and \ref{tab:success} are
  encouraging: the bias on
  $\alpha$ and the coverage probability of confidence intervals
remain good, whichever model is used for estimation. In
  order to confirm this and evaluate the bias on $\alpha$ for larger
  values, another simulation experiment was made. 
For each of the two extreme models exponential and Dirac,
for $\alpha=1,2,\ldots,10$ then $\alpha=10,20,\ldots,100$
and $\rho=0.8,1.0,1.2$, 10000 samples of size 100 were simulated, and
the estimate of $\alpha$ calculated with the other model. It can be
considered that the biases so obtained are an upper bound for the
biases induced by using any of the two extreme cases for an unknown
division time distribution. The relative
bias was calculated as the difference 
between the mean estimate and the true value of $\alpha$, 
divided by the true value of $\alpha$. The results are plotted on 
Figure \ref{fig:ba1100}. For $\alpha\leqslant 3$, the bias is
virtually negligible. For $\alpha\geqslant 4$, estimating as if division
times were constant (red points) induces a positive bias, 
estimating as if they were exponential (green points) induces a negative
bias. 
The relative bias remains smaller than 5\% for $\alpha\leqslant
10$. Notice that in all cases, for any given value of $\alpha$ 
the bias increases with $\rho$.}

{
Having good estimates of the two parameters does not necessarily assure
goodness-of-fit. In another experiment, $10000$ samples of the
$GLD(8,1.2,F)$ were drawn, $F$ being the `realistic' log-normal
distribution. Each sample was adjusted both by the Dirac and
exponential models: $\alpha$ and $\rho$ were estimated for each model and
the goodness-of-fit of the sample with the two adjusted
distributions was tested, using the discrete version of the
Kolmogorov-Smirnov test implemented in the R package 
\verb+dgof+ \cite{ArnoldEmerson11}. The test detected the difference
in about 40\% of the case (39\% of p-values below $0.05$ for the Dirac
model, 43\% for the exponential model). However,
it must be observed that since the data were used to calculate 
the adjusted distribution, the p-values cannot be interpreted as 
if the distribution was independent from the data. More significantly, 
the comparison of Kolmogorov-Smirnov distances showed that the 
Dirac model was a better adjustment
in 67\% of the cases. This is coherent with the results of Table
\ref{tab:bias}.
}
\subsection*{Published data sets}
The simulation study of the previous section indicates that the
estimates of $\alpha$ should be coherent whatever the model, whereas
the exponential model overestimates $\rho$. In order to evaluate the
difference in actual experiments, five sets of published data were used. 
Luria and Delbr\"{u}ck
\cite{LuriaDelbruck43} (Table~2, p.~504) had data
under three different experimental conditions. I have grouped in
sample A experiments numbers 1, 10, 11 and 21b; in sample B
experiments 16 and 17. Data published 
in Boe et al. \cite{Boeetal94}, Rosche and Foster
\cite{RoscheFoster00}, and Zheng \cite{Zheng02} were also used.
For each data set
the  95\% confidence intervals on
$\alpha$ and $\rho$ were computed using the exponential and the Dirac
model. Results are reported in Table \ref{tab:published}. 
The data set from 
\cite{RoscheFoster00} has a high frequency of zeros,
and no jackpot; this explains why $\rho$ cannot be reliably estimated
by the exponential model. The Dirac model gives a more realistic
estimate. 
In all cases, confidence intervals for $\alpha$ are
similar. Confidence intervals on $\rho$ are different, even though
they overlap. As an example, for the Boe et al. data \cite{Boeetal94}, the
estimate of $\rho$ given by the Dirac model is $0.738$; the estimate
given by the exponential model is $0.824$, i. e. 11.6\% larger. That
difference is coherent with what has been observed on simulated
data. Also, the amplitudes of the confidence interval under the
Dirac and exponential models are $0.134$ and $0.172$: the precision
under the Dirac model is better. 

{
The goodness-of-fit was tested for the two models,
using the discrete version of the Kolmogorov-Smirnov test
\cite{ArnoldEmerson11}. The results, reported in Table
\ref{tab:publishedKS}, are not conclusive: both adjustements
are good in all cases. The Dirac model is (slightly) better for
three datasets out of five.
}
\newpage
\section*{Discussion}
{
Dealing with fluctuation analysis experiments and the calculation of
mutation probabilities, three different 
levels must be distinguished: the reality which remains unknown, 
the mathematical model, and the estimation technique.

\noindent
\textit{The unknown reality.} Mutant counts at the end of a
  fluctuation analysis experiment are the result of
\begin{enumerate}
\item a random number of mutations occurring with small probability
  among a large number of cell divisions,
\item the random times during which mutant clones stemming from each
  mutation develop,
\item the number of cells that a clone developing for a given time
  may produce.
\end{enumerate}

\noindent
\textit{The mathematical model.} All models can be interpreted
according to
the same three  points. The first two are
  hardly disputable; the third one is much more controversial.
\begin{enumerate}
\item Due to the law of small numbers, the number of mutations must
  follow a Poisson distribution with expectation $\alpha$, understood
  as the mean number of mutations occurring during the experiment,
  i.e. the product of the individual probability of 
mutation (also called mutation rate) by the final number of cells.
\item The developing time of a random clone has exponential
  distribution with parameter $\rho$, provided the time scale has been
  chosen so that the growth rate of mutants is $1$: $\rho$ is the
  ratio of the growth rate of normal cells to that of mutants, or else
  the relative fitness.
\item The distribution of the number of cells that a clone
  developing for a given time can produce depends on various
  modelling hypotheses, such as:
\begin{itemize}
\item if a mutation occurs during a division, 
only one of the two daughter cells is a mutant
\item mutant clones develop forever as mutants (no back mutation)
\item no cell dies before dividing
\item the division times are independent and identically distributed
\item the distribution of division times is exponential
\end{itemize}
\end{enumerate}
Since the early forties (and maybe even before: see \cite{Sarkar91}),
mathematicians have struggled to propose sets of modelling hypotheses
that allowed explicit computations of probabilistic distributions.
Since Lea and Coulson \cite{LeaCoulson49}, Bartlett
\cite[p.~37]{Armitage52}), and Haldane \cite{Sarkar91,Zheng07}, the
first four of the above hypotheses have been widely agreed upon. As
for the distribution of division times, the exponential model that
leads to the classical Luria-Delbr\"uck distribution has largely
prevailed \cite{Zheng99,Zheng10}, though constant division times
have also been considered \cite{Sarkar91,Zheng07}. At first, only the
case were normal cells and mutants had the same growth rate ($\rho=1$
in our notations) was studied. But soon, with 
Mandelbrot \cite{Mandelbrot74} and Bartlett 
\cite[section 4.31]{Bartlett78}, the model was generalized to
differential growth rates
\cite{Koch82,Jones94,JaegerSarkar95,Zheng02,Zheng05}.
Strangely enough, whereas the Poisson approximation (point 1. above) has been
considered an obvious fact since Luria and Delbr\"uck
\cite{LuriaDelbruck43}, the exponential distribution of
development times (point 2.) has remained unnoticed, even
though it was known as a basic fact of branching process theory at
least since the sixties \cite{Kendall66}. It was remarked in
\cite{HamonYcart12}, and leads not only to a much simpler derivation of
closed mathematical formulas, but also to simple and efficient
simulation algorithms.

A distinctive hypothesis of the model considered here (as in most
previous  works),
is that cells can only divide and never
die. A model taking cell deaths into account was described in
\cite{Ycart13a}, and an estimation procedure was proposed. In practice,
the proportion of deaths is known to be rather low
\cite{Stewartetal05,Fontaineetal08}. As shown in
\cite{Ycart13a}, neglecting cell deaths underestimates
$\alpha$ and $\rho$. 
Another dubious hypothesis of the models
considered so far is the independence of individual division times.
The independence hypothesis was
questioned very early \cite{Kendall52a}. Indeed, actual
division time data show two types of correlation \cite{Wangetal10}: 
between the division times of a
mother cell and its two daughters, and between the two sisters
conditioning on the mother. It was remarked long ago
by Powell \cite{Powell56} (see also \cite{CrumpMode69a,Harvey72}) that
sister-correlations do not influence 
exponential growth. The effect of mother-correlation on growth rates
was discussed by Harvey in \cite{Harvey72}. Its influence on the
estimation of parameters in fluctuation analysis will be the object
of future work.

\noindent
\textit{The estimation technique.}
From Luria and Delbr\"uck \cite{LuriaDelbruck43}, the mean number of mutations
$\alpha$ has been
the parameter of interest, whereas the relative fitness
$\rho$ was regarded at best as a nuisance parameter, or very often
taken as fixed: $\rho=1$
\cite{Stewartetal90,Angerer01b,Foster06}. Indeed,  
the relative fitness can be independently estimated,
by separately growing clones of mutants and normal cells, and
calculating their growth rates \cite{KimmelAxelrod02}. 
If this has been done, then $\rho$ can
be considered as known, which leads to a better estimation of
$\alpha$, as pointed out  in \cite{HamonYcart12}. Yet $\rho$ is rarely
known in practice. Its independent calculation may be
difficult in some cases (in vivo experiments for instance). 
Considering
differential growth rates is necessary, as pointed out by 
several authors
\cite{Koch82,Jones94,JaegerSarkar95,Zheng02,Zheng05,HamonYcart12};
however, many studies are still being made using   
the LD$(\alpha,1)$ without questionning the equal rate hypothesis
(e.g.  \cite{Wuetal09,Jeanetal10}).

Once a mathematical model has been chosen, many estimation procedures
for $\alpha$ and/or $\rho$ are available \cite{Foster06,HamonYcart12}. 
As in any parametric estimation problem, the questions are:
\begin{itemize}
\item are estimates unbiased?
\item can confidence intervals be computed?
\item is the mean squared error minimal?
\end{itemize}
Only three methods answer positively the first two questions:
the $p_0$-method
\cite{LuriaDelbruck43,Foster06}, the Maximum
Likelihood (ML) method \cite{Maetal92,Jonesetal93,Zheng02,Zheng05}, 
and the Generating Function (GF) method \cite{HamonYcart12}.
As in many other estimation problems, the best method in terms of
mean squared error is the ML method. As was shown in
\cite{HamonYcart12}, the $p_0$-method performs well
for small values of $\alpha$. The GF method is nearly as precise as the ML,
with a much broader range of applicability, and virtually
null computing time. 

To go further, three more criteria must be added:
\begin{itemize}
\item to how many models can the procedure be applied?
\item can it work on a wide enough range of values of $\alpha$ and $\rho$?
\item is it robust to variations of modelling hypotheses, or else how
  much bias estimating with a wrong model does induce? 
\end{itemize}
As far as the first and third questions are concerned, the winner
is the $p_0$-method: the distribution of the estimator is easily
computed under any model, and the result does not depend on any
hypothesis, except the fact that cells always divide and never die
(see \cite{Ycart13a} for an alternative in the case of cell deaths).
However, it relies upon a positive number of zeros in the sample, and is
therefore limited to relatively small values of $\alpha$ (smaller than $2$ in
practice). Such a limitation is not statistically acceptable.

Regarding the first question, 
the ML procedure can be applied if the probabilities of mutant
counts can be computed as a function of the parameters. 
This is the case for only two distributions
so far: the classical $LD(\alpha,\rho)$ (independent exponential
division times, no deaths), and the $GLD(\alpha,\rho,F_{\mathrm{dir}})$
(constant division times, no deaths). The GF procedure can be applied to any
$GLD(\alpha,\rho,F)$, provided the distribution $F$ has been previously
estimated. It was
applied to a cell-death model in \cite{Ycart13a}. 
Actually the Monte-Carlo algorithm proposed here can be
used for any model, as soon as clones can be simulated. 
If the distribution of division times is unknown, any 
one of the two models above (exponential or constant division times)
can be chosen.

The second question has been discussed in \cite{HamonYcart12}. 
Even with a very careful algorithmic implementation
\cite{Zheng02,Halletal09}, 
the ML method can compute estimates only for samples in which the
maximal value does not exceed a certain limit.
Yet a crucial feature of mutant counts is the appearance
of jackpots, i.e. unusually large values. For the ML method to be
applied, the highest jackpots must be levelled out, which induces 
a systematic bias both on $\alpha$ and $\rho$. 
This explains why the ML method can be used only
when large jackpots are very unlikely, or else if $\alpha$ is small
enough  and $\rho$ large enough; as an indicative range of values,
$\alpha<10$ and $\rho>0.5$ can be considered. Admittedly, current
experiments stay within that
range, but for how long?  

Regarding the third question, estimating parameters with a wrong model
can be expected to induce some
bias, whichever estimation method is used. As shown in
\cite{HamonYcart12}, the GF and ML methods output very similar results
(when both can be used). So the conclusions of the `Results' section 
would hold as well for ML estimates. The main question was to evaluate
which bias could be expected from using either the Dirac or the classical
exponential model, when data were simulated using a more realistic model.
The estimation of $\alpha$ can be expected 
to be robust for low values,
because when $\alpha$ is small, the information is
concentrated on the first value $p_0=\mathrm{e}^{-\alpha}$ that depends
only on $\alpha$. The surprise was that it is still robust up to
$\alpha=10$, when $p_0$ is very small and the
$p_0$-method cannot be used (Figure
\ref{fig:ba1100}, left panel). 
For very large values of $\alpha$, both models induce a bias on
$\alpha$, positive for the Dirac model, negative for the exponential
model (Figure \ref{fig:ba1100}, right panel). 
The estimation of $\rho$ is more sensitive to the
model: estimating with the exponential model induces a
positive bias; using the Dirac model reduces the bias
(Figure \ref{fig:boxplots} and Table
\ref{tab:bias}). 
}

\section*{Models and Methods}
\subsection*{Division time distributions and growth rates}
\label{gr}
In this section, the probabilistic model of cell division and
mutations is described,
the relation between division times and growth rates is precised, and
the goodness-of-fit of Kelly and Rahn's data \cite{KellyRahn32} with
three families of distributions is detailed.

In Kendall's notation \cite{Kendall52}, 
the model considered here is G/G/0:
\begin{itemize}
\item at time $0$ a homogeneous culture of $n$ normal cells is given;
\item the division time of any normal cell is a random variable with
  distribution function $G$;
\item when the division of a normal cell occurs, it is replaced by:
\begin{itemize}
\item one normal and one mutant cell with probability $p$,
\item two normal cells with probability $1-p$;
\end{itemize}
\item the division time of any mutant cell is a random variable with
  distribution function $F$;
\item when the division of a mutant cell occurs, it is replaced by two
  mutant cells;
\item all random variables and events (division times and mutations) 
are mutually independent.
\end{itemize}
The probabilistic results used here come from the theory of
continuous time branching processes: see \cite{Harris63,AthreyaNey72}.
To a distribution of division times corresponds an exponential
growth rate for the corresponding clones: the growth rate of a
clone with binary divisions is
the point at which the Laplace transform of division times 
equals $1/2$. If all division times are multiplied by a constant,
the growth rate is divided by the same constant. Therefore scaling a
distribution to have unit growth rate amounts to multiplying all
division times by the initial growth rate. Here
I assume that the time scale has been chosen so that the growth rate
of normal cells is the relative fitness $\rho$, and the growth rate of
mutants is $1$:
$$
\int_0^{+\infty} \mathrm{e}^{-\rho t}\,\mathrm{d} G(t) =
\int_0^{+\infty} \mathrm{e}^{- t}\,\mathrm{d} F(t) =\frac{1}{2}\;.
$$
Two particular cases will be seen as extreme values for the
distribution $F$: exponential and Dirac distributions.
$$
F_{\mathrm{exp}}(t)=1-\mathrm{e}^{-t}
\;;\quad
F_{\mathrm{dir}}(t) = \mathbb{I}_{[\log(2),+\infty)}(t)\;,
$$
where $\mathbb{I}$ denotes the indicator function of an interval ($1$ or $0$
according to whether the variable is in the interval or not). 
These distributions have coefficients of variation equal to $1$ and $0$
respectively. The coefficients of variation observed in experiments
are of order $0.2$ \cite{Nivenetal08}. 
I have chosen three families of
distributions to illustrate my results: 
Gamma, Log-normal and Inverse Gaussian. All three have
the property to be invariant through scaling. For instance, 
if $X$ has Gamma $GA(a,\lambda)$
distribution, then $sX$ has $GA(a,\lambda/s)$ distribution; similar
relations hold for the two other families.
The probability
distribution functions, Laplace transforms,
and scaling parameters are given in Table 
\ref{tab:GALNIG}. As many other families of distributions, these three
encompass the two extremes of exponential and Dirac distributions 
as limit cases and interpolate between them. 
This is illustrated by Figure \ref{fig:GALNIG} where 20 densities of unit
growth rate distributions are plotted for each family.

In order to get one realistic distribution per family, 
the historical observations of
Kelly and Rahn on Bacterium aerogenes \cite[Table~2,
p.~149]{KellyRahn32}
were adjusted.
A maximum likelihood estimation of the parameters
on the data led to one particular distribution in each family, 
that was rescaled to unit growth rate. Figure \ref{fig:BA} illustrates
the fit. On the left panel, the histogram and the 3 densities are
superposed; the right panel displays the corresponding densities
after scaling to unit growth rate. Table \ref{tab:BA} gives the
parameters of the three densities, together with the p-values of the
Kolmogorov-Smirnov and Anderson-Darling goodness-of-fit tests: all
three fits turn out to be satisfactory.
\subsection*{Generalized Luria-Delbr\"{u}ck distributions}
\label{gld}
Consider an initial (large) number $n$ of normal cells. Assume
that the mutation probability $p$ is small, that the time $t$
at which mutants are counted is large, and that the asymptotics are
such that the expected number of mutations $\alpha$ before
time $t$ is non null and finite. 
Using general results of branching process theory  
\cite[Chap. VI]{Harris63} and \cite[Chap. IV]{AthreyaNey72}, it
can be proved that the total number of mutants at time $t$ approximately
follows an integer valued distribution, whose probability 
generating function (PGF) is given by:
\begin{equation}
\label{ldg}
g_{\alpha,\rho}(z) = \exp(\alpha(h_{\rho}(z)-1))\;,
\end{equation}
with:
\begin{equation}
\label{yg}
h_{\rho}(z)=\int_0^{+\infty} \psi(z,t) \,\rho\mathrm{e}^{-\rho t}\,\mathrm{d} t\;,
\end{equation}
where $\psi(z,t)$ is the PGF of $M(t)$, i.e. the number of cells at time $t$ 
in a mutant clone, starting from
one single cell at time $0$.
\begin{equation}
\label{psi}
\psi(z,t)= \mathbb{E}[z^{M(t)}]\;,
\end{equation}
where $\mathbb{E}$ denotes mathematical expectation. The
  explicit expressions (\ref{ldg}) and (\ref{yg}) are quite general,
  and do not depend on any modelling assumption apart from exponential
  proliferation. If the individual division times of mutants are
  supposed to be independent with common distribution $F$, then 
the function $\psi(z,t)$ is uniquely defined in terms of $F$.

The interpretation of (\ref{ldg}) and (\ref{yg}) is quite simple, and
can be separated into the following two arguments. 
\begin{itemize}
\item[(\ref{ldg})]
The number of mutations converges in distribution to 
the Poisson distribution with parameter $\alpha$ (this remark had
already been made by Luria and Delbr\"{u}ck  
\cite[p.~499]{LuriaDelbruck43}). From each mutation stems
a mutant clone that develops at final time $T$ into a random number of
mutants, each with PGF $h_\rho$. A random number of such clones must be
added: the result is a Poisson sum of independent random variables
with PGF $h_\rho$. This yields equation (\ref{ldg}).
\item[(\ref{yg})]
Any given mutation happens at some
division instant chosen at random (i.e. uniformly distributed) 
among all division instants. Due to exponential growth, division 
instants are more concentrated near the end of the observation
interval. It can be proved that the difference between the
final time and a
randomly chosen division instant, i.e. the developing time of a
typical mutant clone, is exponentially distributed with parameter
$\rho$. Therefore the size at final time of a typical mutant clone is
an exponential mixture of sizes at fixed time $t$. Hence equation
(\ref{yg}). 
\end{itemize}
Precise mathematical statements and proofs of the asymptotics described
above have been given in \cite{HamonYcart12}, and will not be
reproduced here. I propose to name \emph{Generalized
Luria-Delbr\"{u}ck} distribution with parameters $\alpha$, $\rho$,
and $F$ and denote by $GLD(\alpha,\rho,F)$, the probability distribution on the
set of integers whose PGF $g_{\alpha,\rho}$ is defined by (\ref{ldg})
and (\ref{yg}). Observe that it depends on the division time
distribution of normal cells $G$ only through the growth rate $\rho$,
whereas it does depend on the actual division time distribution $F$ of
mutant cells. The particular case $GLD(\alpha,\rho,F_{\mathrm{exp}})$
is the classical Luria-Delbr\"{u}ck distribution $LD(\alpha,\rho)$.
In that case,
$$
F_{\mathrm{exp}}(t)=1-\mathrm{e}^{-t}\;,
$$
and 
\begin{equation}
\label{hexpo}
h^{\mathrm{exp}}_\rho(z)=\int_0^{+\infty} \frac{z \mathrm{e}^{-t}}{1-z+z\mathrm{e}^{-t}}
\,\rho\mathrm{e}^{-\rho t}\,\mathrm{d} t\;. 
\end{equation}
The exponential case has been known for a long time: see Zheng
\cite{Zheng99,Zheng10} for historical accounts. As shown in
\cite{HamonYcart12},
formula (\ref{hexpo})
comes from the fact that the size of a mutant clone at time $t$ follows
the geometric distribution with parameter $\mathrm{e}^{-t}$, a fact
already pointed out by Yule 
\cite[p.~35]{Yule25} (see also \cite[p.~109]{AthreyaNey72}). 
It turns out that 
explicit expressions of $\psi(z,s)$, $h_\rho(z)$, and $g_{\alpha,\rho}$
can also be given in the case where division times are constant, which
is the object of the next section. 

\subsection*{Constant division times}
Here it is assumed that division times of mutants are constant, i.e. 
$F$ is the Dirac distribution at $\log(2)$ (to ensure unit
growth rate).
$$
F_{\mathrm{dir}}(t)=\mathbb{I}_{[\log(2),+\infty)}(t)\;.
$$
Thus the generalized Luria-Delbr\"uck distribution 
$GLD(\alpha,\rho,F_{\mathrm{dir}})$ is considered.
The idea can be traced back to Haldane who used it to propose an
approximation heuristics for calculating the probabilities of mutant
counts: see Sarkar \cite{Sarkar91} and Zheng \cite{Zheng07} (actually,
Haldane's model can be related to the particular case
$GLD(\alpha,1,F_{\mathrm{dir}}$).

For the $GLD(\alpha,\rho,F_{\mathrm{dir}})$, formula (\ref{yg}) becomes:
\begin{equation}
\label{hconst}
h^{\mathrm{dir}}_\rho(z)=(1-2^{-\rho})\sum_{n=0}^{+\infty} 2^{-n\rho} z^{2^n}\;. 
\end{equation}
To the best of my knowledge (\ref{hconst}) is new. Here is how
it is derived. 
With constant division times, say all equal to $a$, the population doubles at
multiples of $a$. Hence the exponential growth rate is
$\log(2)/a=1$, therefore 
$a=\log(2)$. Between instants $na$ and $(n+1)a$, there are $2^n$ cells in the
clone. Hence the generating function at time $s$:
$$
\psi^{\mathrm{dir}}(z,s) = \sum_{n=0}^{+\infty} z^{2^n} \,\mathbb{I}_{[na,(n+1)a)}(s)\;.
$$ 
Integrating against the exponential distribution with parameter
$\rho$ gives:
$$
h^{\mathrm{dir}}_\rho(z) =  \sum_{n=0}^{+\infty} z^{2^n}\,
\mathrm{e}^{-na\rho}(1-\mathrm{e}^{-a\rho})\;,
$$
hence (\ref{hconst}), since $\mathrm{e}^{a}=2$.

Not only the PGF, but also the probabilities of the 
$GLD(\alpha,\rho,F_{\mathrm{dir}})$ can be easily computed. 
Indeed, let $(p_k)_{k\geqslant 1}$ denote the probabilities of
the distribution with PGF $h^{\mathrm{dir}}_\rho$:
$$
p_k=(1-2^{-\rho}) 2^{-n\rho} \mbox{ if } k=2^n\,,\; 0 \mbox{ else.}
$$ 
Let $(q_k)_{k\geqslant 0}$ be the probabilities  of the 
$GLD(\alpha,\rho,F_{\mathrm{dir}})$. They can be computed by
the following well known recursive
formula, easily deduced from the probability generating
function (\ref{ldg})  
(see \cite{Pakes93} and references therein):
\begin{equation}
\label{algo}
q_0=\mathrm{e}^{-\alpha}
\;\mbox{and for $k\geqslant 1$, }
q_k = \frac{\alpha}{k} \sum_{i=1}^k ip_i q_{k-i}\;.
\end{equation}
The algorithm has been encoded in the R script available online: the
probabilities, cumulated distribution function and quantile function
of the $GLD(\alpha,\rho,F_{\mathrm{dir}})$ are provided.
The log-likelihood and its derivatives
with respect to the parameters also have explicit algorithms, almost
identical to those implemented for the $LD(\alpha,\rho)$ by Zheng
\cite{Zheng05}. The conclusion is that the estimation of $\alpha$ and
$\rho$ can be conducted for the $GLD(\alpha,\rho,F_{\mathrm{dir}})$
exactly as for the $LD(\alpha,\rho)$, either by the classical Maximum
Likelihood method \cite{Zheng05} or by the Generating function method
\cite{HamonYcart12}. The algorithms are even faster and numerically more
stable in the constant division time model.
\subsection*{General division times}
No distribution $F$ other than $F_{\mathrm{exp}}$ and
$F_{\mathrm{dir}}$ leads to such closed
expressions as (\ref{hexpo}) and (\ref{hconst}).
However, it is possible to compute numerically
$h_\rho(z)$ for any $F$, using a Monte-Carlo algorithm that 
will now be described. If a division time distribution is given, 
sequences of independent division times can be simulated at will. 
From such a sequence, a
clone can be simulated up to any arbitrary time, outputing the number of
cells as a function of time. That function of time is encoded 
by the sequence of instants at which the function increases by $1$,
i.e. when divisions occur. Choose a value
$\rho_{\min}$, such that 
any subsequent evaluation of $h_\rho(z)$ will be made for values of
$\rho$ larger than $\rho_{\min}$. In simulations, I have chosen
$\rho_{\min}=0.8$, but this value could be adjusted.
Let $T_1,\ldots,T_k$ be $k$ independent instants, simulated
according to the exponential distribution with parameter
$\rho_{\min}$. A crucial observation is that if $T_h$ is exponentially
distributed with parameter $\rho_{\min}$, then for any $\rho\geqslant
\rho_{\min}$, $\frac{\rho_{\min}}{\rho} T_h\leqslant T_h$ is
exponentially distributed with 
parameter $\rho$.
For $h=1,\ldots,k$, denote by $N_h(t)$ the number of living
cells at time $t$ in a random clone, starting from a single mutant
cell at time $0$, simulated up to time $T_h$. For any $\rho\geqslant
\rho_{\min}$, and any $z\in [0,1]$,  consider:
$$
\hat{h}_\rho(z) = \frac{1}{k}\sum_{h=1}^k z^{N_h(T_h\rho_{\min}/\rho)}\;.
$$
By the law of large numbers, as $k$ tends to infinity,
$\hat{h}_\rho(z)$ converges  to $h_\rho(z)$.
The central limit theorem yields a precision of order $1/\sqrt{k}$ on
the result. In  simulations (in particular to compute
  the curves of Figure
  \ref{fig:hr}), $k$ has been fixed to
$10^{5}$. Observe that
the (time consuming) simulation of the $k$ clones needs to be done only
once: from there, all subsequent evaluations of $\hat{h}_\rho(z)$ will
be deduced.

As will be seen in the next section, the estimation 
of $\alpha$ and $\rho$ and the computation of their confidence intervals
require repeated evaluations of the derivative in $\rho$ of $h_\rho(z)$.  
Using the procedure above to evaluate that derivative 
by finite differences would lead to quite unprecise
results. Another procedure,
similar to the previous one, is proposed instead. 
The derivative in $\rho$ of $h_\rho(z)$ is:
\begin{eqnarray*}
\displaystyle{\frac{\partial h_\rho(z)}{\partial \rho}}&=&
\displaystyle{\int_0^{+\infty} \psi(z,s)\,\mathrm{e}^{-\rho s}\,\mathrm{d} s
- \rho \int_0^{+\infty} \psi(z,s)\,s\mathrm{e}^{-\rho s}\,\mathrm{d}
  s}\\[2ex]
&=&\displaystyle{\frac{1}{\rho}\,h_\rho(z) -\frac{1}{\rho}\,\tilde{h}_\rho(z)}\;,
\end{eqnarray*}
with:
$$
\tilde{h}_\rho(z) = \int_0^{+\infty} \psi(z,s)\,\rho^2s \mathrm{e}^{-\rho
  s}\,\mathrm{d} s\;.
$$
Now $\rho^2 s\mathrm{e}^{-\rho s}$ is the density of the Gamma distribution
$GA(2,\rho)$ (sum of two independent
exponentially distributed random variables). Therefore
$\tilde{h}_\rho(z)$ is the PGF of the number of cells in a clone
starting from a single mutant cell at time $0$, observed up to an
independent, Gamma distributed random time. 
Let $\tilde{T}_1,\ldots,\tilde{T}_k$ be $k$ independent instants, simulated
according to the Gamma distribution with parameters $2$ and $\rho_{\min}$.
For $h=1,\ldots,k$, denote by $\tilde{N}_h(t)$ the number of living
cells at time $t$ in a random clone, starting from a single mutant
cell at time $0$, simulated up to time $\tilde{T}_h$. For any $\rho\geqslant
\rho_{\min}$, and any $z\in [0,1]$,  consider:
$$
\hat{\tilde{h}}_\rho(z) = \frac{1}{k}\sum_{h=1}^k 
z^{\tilde{N}_h(\tilde{T}_h\rho_{\min}/\rho)}\;.
$$
By the law of large numbers, as $k$ tends to infinity,
$\hat{\tilde{h}}_\rho(z)$ converges  to $\tilde{h}_\rho(z)$.

Further savings in computer time can be obtained by the following
remark. Let $\tilde{T}$ follow the Gamma distribution with parameters
$2$ and  $\rho$. Let $U$ be another random variable, independent
from $T$, uniformly distributed on the interval $[0,1]$. Then $UT$
follows the exponential distribution with rate $\rho$. Therefore 
the same $k$ clones, simulated up to Gamma distributed
instants, can be used to estimate the values of both $h_\rho(z)$ and
$\tilde{h}_\rho(z)$.
\subsection*{Generating function estimators} 
\label{epgf}
The main goal of fluctuation analysis is to estimate the mutation
probability $p$, from a sample of mutant counts. If an estimate
of the mean number of mutations $\alpha$ has been calculated, then an
estimate of $p$ can be deduced, dividing by the final number of
cells: the parameter of main interest is $\alpha$.
Many methods of estimation for $\alpha$ have been proposed: see
\cite{Foster06}. The simplest consists in estimating the probability
of observing no mutant: $\mathrm{e}^{-\alpha}$; this is the original
method used by Luria and Delbr\"{u}ck 
\cite{LuriaDelbruck43}, and is usually referred to as
``$p_0$-method''. Observe that the result does
not depend on $\rho$, nor on $F$. {Therefore the
  $p_0$-method is completely independent from any modelling
  hypothesis. It that sense it is the most robust of all methods}.
However,  the $p_0$-method can be used only
if $\alpha$ is small enough {(so a sizeable number of tubes do not
contain any mutant).} As explained in \cite{HamonYcart12} and in the
discussion section, such a
limitation cannot be accepted.

Apart from the $p_0$-method,  any other consistent
estimator of $\alpha$ must depend on 
the value of $\rho$ and on the mutant division time distribution
$F$. Maximum Likelihood is usually considered 
the  best  estimation method in a
parametric inference problem. For the estimation of the
parameters $\alpha$ and $\rho$ of the classical $LD(\alpha,\rho)$, it
has been recommended by several authors: 
\cite{Maetal92,Jonesetal93,Zheng02,Zheng05}. In
\cite{HamonYcart12}, its limitations were pointed out, and an
alternative procedure, based on the empirical probability generating
function (EPGF), was proposed. It turns out
that the EPGF method can be adapted to the general case of the
$GLD(\alpha,\rho,F)$, whereas the Maximum Likelihood cannot. 
It only relies upon the numerical 
evaluations of $h_\rho(z)$ and its derivative
in $\rho$. For the $LD(\alpha,\rho)$ and
  $GMD(\alpha,\rho,F_{\mathrm{dir}})$ explicit formulas are available,
  for the other cases a Monte-Carlo algorithm was 
described in the previous section.
The procedure is described below, and the reader is refered 
to the R functions that have been made available online 
for implementation details:  they include estimation, 
confidence intervals, and hypothesis testing.

Let $(X_1,\ldots,X_n)$ be a sample of independent random variables, each
with $GLD(\alpha,\rho,F)$ distribution. 
Recall the probability generating function of the $GLD(\alpha,\rho,F)$:
$$
g_{\alpha,\rho}(z)=\exp(\alpha (h_\rho(z)-1))\;,
$$ 
with:
$$
h_\rho(z) = \int_0^{+\infty}\psi(z,t)\,\rho\mathrm{e}^{-\rho t}\,\mathrm{d} t\;.
$$
Define the
empirical probability generating function (EPGF) $\hat{g}_n(z)$ as:
$$
\hat{g}_n(z) = \frac{1}{n} \sum_{i=1}^n z^{X_i}\;.
$$
The random variables $z^{X_i}$ are bounded and mutually independent:
by the law of large numbers, $\hat{g}_n(z)$ is a
consistent estimator of $g_{\alpha,\rho}(z)$, for any $z$ in
$[0,1]$.
For $0<z_1<z_2<1$, consider the following
ratio:
\begin{equation}
\label{deff}
f_{z_1,z_2}(\rho)=\frac{h_\rho(z_1)-1}{h_\rho(z_2)-1}
=\frac{\log g_{\alpha,\rho}(z_1)}{\log g_{\alpha,\rho}(z_2)}
\;.
\end{equation}
The function that maps $\rho$ onto 
$y=f_{z_1,z_2}(\rho)$ is continuous and strictly monotone, hence
one-to-one. Therefore the inverse, that maps $y$ onto
$\rho=f^{-1}_{z_1,z_2}(y)$, is well
defined.
For $0<z_1<z_2<1$, let $\hat{y}_n(z_1,z_2)$ denote the following log-ratio.
$$
\hat{y}_n(z_1,z_2) = 
\frac{\log(\hat{g}_n(z_1))}{\log(\hat{g}_n(z_2))}\;.
$$
An estimator of $\rho$ is obtained by:
$$
\hat{\rho}_n(z_1,z_2) = f^{-1}_{z_1,z_2}(\hat{y}_n)
$$
Then an estimator of $\alpha$ by:
$$
\hat{\alpha}_n(z_1,z_2,z_3) = \frac{\log(\hat{g}_n(z_3))}
{h_{\hat{\rho}_n(z_1,z_2)}(z_3)-1}\;,
$$
where $z_3\in(0\,;1)$ is a new control, possibly different from
$z_1$ and $z_2$. 
Observe that $\hat{\alpha}_n(z_1,z_2,z_3)$ depends on
$\hat{\rho}_n(z_1,z_2)$, whereas
$\hat{\rho}_n(z_1,z_2)$ only depends on the arbitrary choice of the
couple $(z_1,z_2)$. 
They will be referred to as generating function (GF) estimators.
The strong consistence and asymptotic variance of the GF estimators
were studied in \cite{HamonYcart12}, and
mathematical details will not be reproduced 
here. In particular, Proposition 4.1 of that
reference gives the explicit form of the asymptotic covariance matrix,
upon which inference procedures are based (confidence intervals and
hypothesis testing). The asymptotic covariance matrix has been encoded
in the $R$ functions made available online; it expresses in terms of:
\begin{itemize}
\item the PGF $g_{\alpha,\rho}$ evaluated at $z_1$, $z_2$, $z_3$, and
  their products two by two,
\item the PGF $h_\rho$ evaluated at $z_1$, $z_2$, $z_3$,
\item the derivative in $\rho$ of $h_\rho$, 
evaluated at $z_1$, $z_2$, $z_3$.
\end{itemize}

The GF estimators 
depend on the three arbitrary values of $z_1$, $z_2$ and $z_3$. Another
tuning parameter has to be added. In the $GLD(\alpha,\rho,F)$
the parameter $\rho$,
determines the size and frequency of much larger
values than usual (called ``jackpots'' in \cite{LuriaDelbruck43}). 
For $\rho<1$, some very large values can be
obtained, even for a small $\alpha$. Using the empirical probability
generating function  is a simple way to damp down jackpots, 
and get robust estimates. The
variable $z$ can be seen as a tuning parameter for the damping. 
At the limit case $z=0$, 
$\hat{g}_n(0)$ is simply the frequency of null values, 
and $\hat{\alpha}_n(0)=-\log(\hat{g}_n(0))$ is the
so called $p_0$-estimator of $\alpha$, already proposed in 
\cite{LuriaDelbruck43} (it does not depend on
$\rho$ nor $F$). For $z_1=0.1$,
only small observations will be taken into account, whereas for
$z_2=0.9$, much larger values will influence the sum. Thus the
empirical probability generating function
damps down jackpots in a differential way according to $z_1$ and
$z_2$. Choosing $z_1=0.1$ and 
$z_2=0.9$ will contrast small values compared to jackpots,
which explains why $\hat{\rho}_n$ can efficiently estimate
$\rho$ for small $\alpha$'s. 
However, for large values of $\alpha$ (say $\alpha>5$), even
$z_2=0.9$ will output very small values, below the machine
precision. This will make the estimates numerically unstable. A
natural way to stabilize them is to rescale the sample, 
dividing all values by a common factor $b$. This amounts to replacing
$z$ by $z^{1/b}$ in the definition of $\hat{g}_n(z)$:
$$
\frac{1}{n}\sum_{i=1}^n z^{X_i/b} =
\frac{1}{n}\sum_{i=1}^n (z^{1/b})^{X_i}=\hat{g}_{n}(z^{1/b})\;. 
$$
I proposed to set $b$ to the $q$-th quantile of the
sample, where $q$ is another control. Based on simulation evidence,  
my best compromise is $z_1=0.1$, $z_2=0.9$, $z_3=0.8$, $q=0.1$.
In the implementation of the GF estimators, the scaling
factor $b$ is set to the $q$-th quantile of the
sample, and all data are divided by that scaling factor (which
amounts to replacing $z_1,z_2,z_3$ by $z_1^{1/b},z_2^{1/b},z_3^{1/b}$). The
estimators $\hat{\alpha}_n$ and $\hat{\rho}_n$ are computed with these
values.

The GF estimators crucially rely upon the inverse of the function $f$,
defined by (\ref{deff}). Figure \ref{fig:hr} shows variations of $f$
according to the underlying model. On that figure, plots of $f$ for
the exponential and Dirac model have been represented, together with
plots of $f$ for the 3 distributions determined by
fitting actual data. The curves corresponding to realistic
distributions are close together, and closer to the Dirac case than to
the exponential case. From this graphics, it can be anticipated that
estimating $\rho$ with the classical Luria-Delbr\"{u}ck model 
induces a positive bias; this was indeed observed on simulations. Also
the fact that the slope of the curve corresponding to the exponential
model is smaller  explains why the
precision on $\rho$ obtained by the classical method is worse. 
 
\section*{Acknowledgments}
Thanks to Jo\"el Gaff\'e, Agn\`es Hamon,
Alain Le Breton, and Dominique Schneider for helpful 
and pleasant discussions. {I am indebted to the anonymous
  reviewers for important suggestions.}

\bibliography{/home/ycart/recherche/IS/IS.bib}
\newpage
\section*{Figure Legends}

\begin{figure}[!ht]
\begin{center}
\begin{tabular}{cc}
\includegraphics[width=0.45\linewidth]{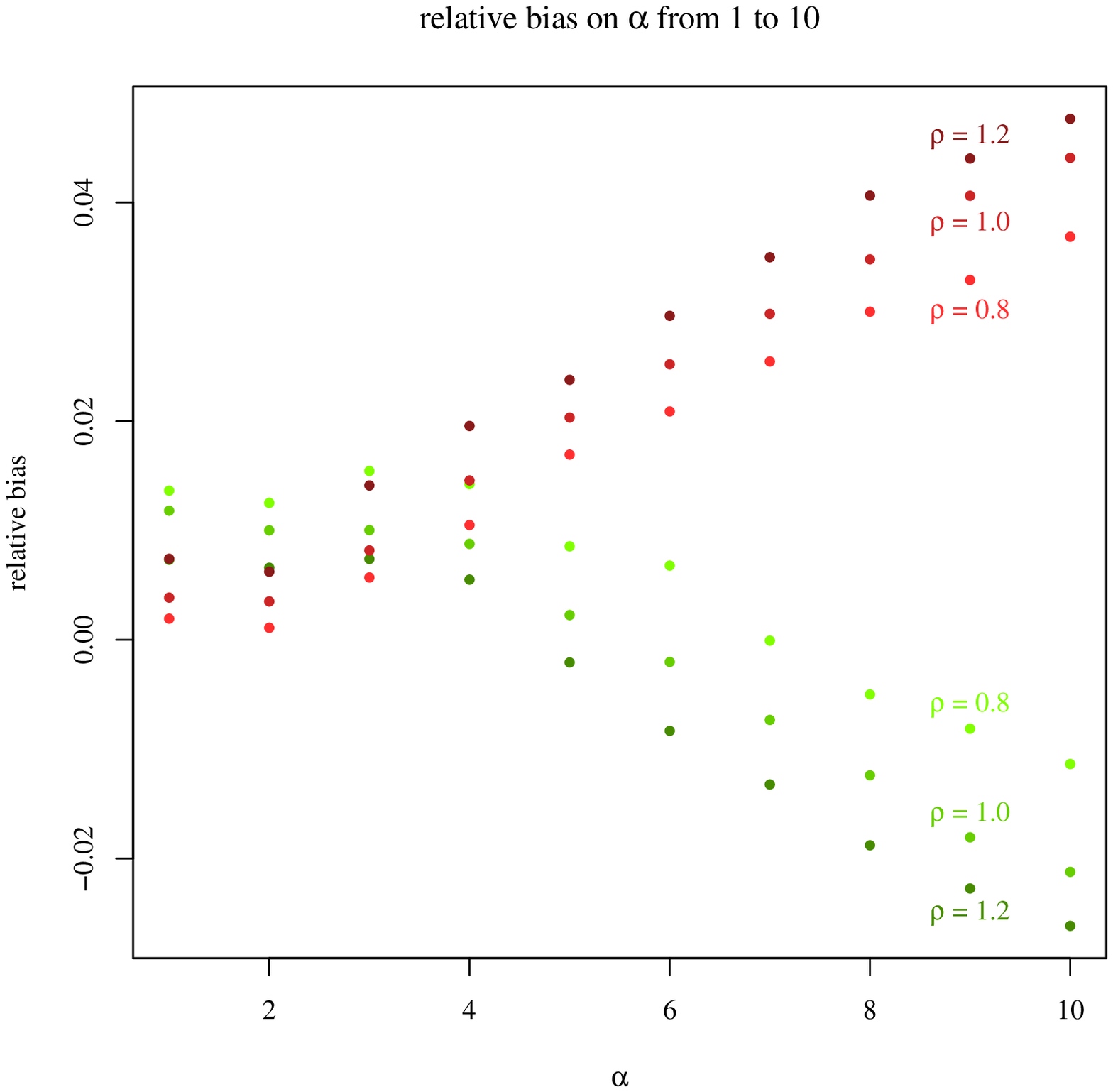}&
\includegraphics[width=0.45\linewidth]{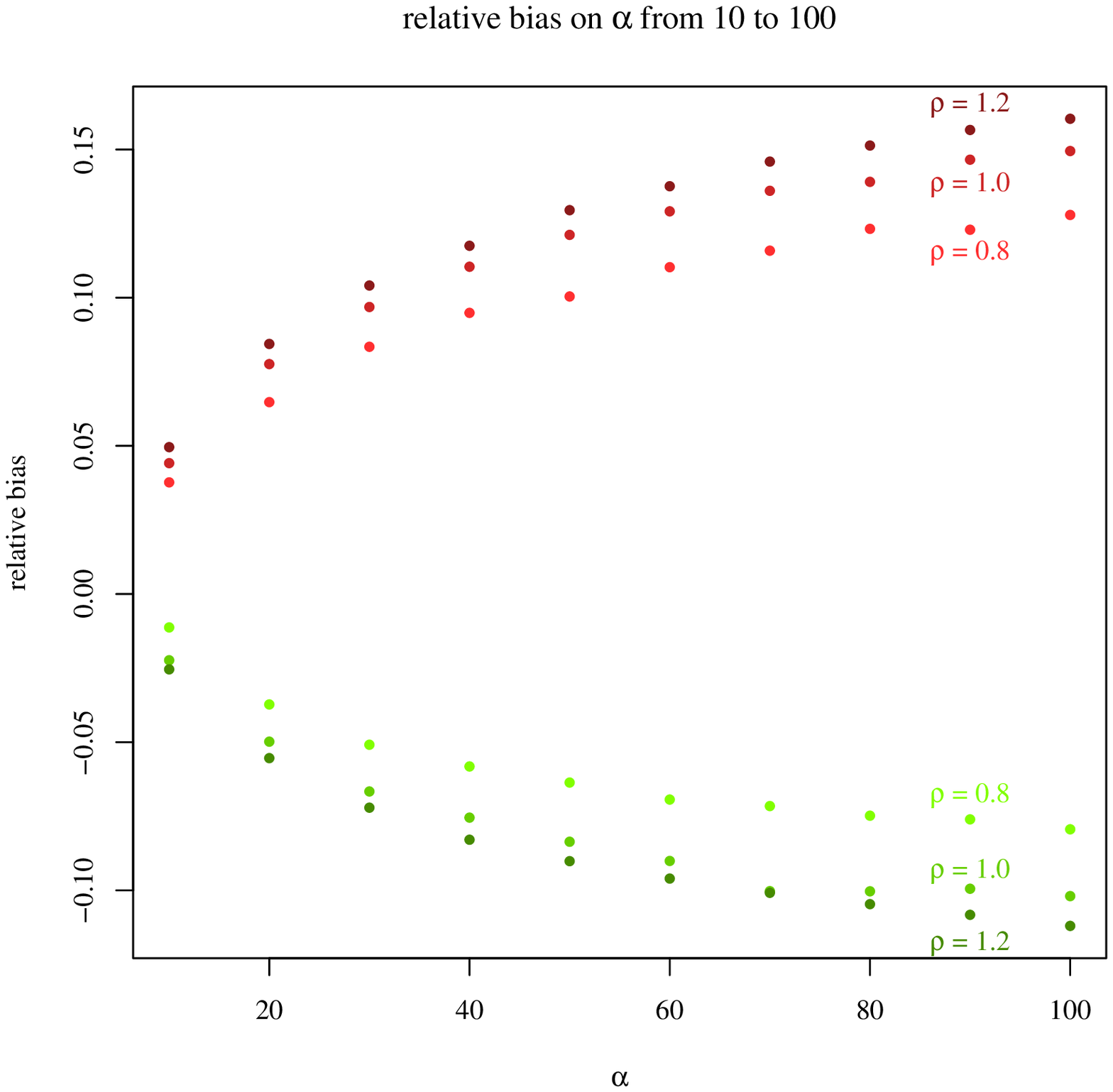}
\end{tabular}
\end{center}

\caption{{{\bf Relative bias on $\alpha$ between the
  exponential and the Dirac models.} Ten thousand samples of size 100
were simulated for the $LD(\alpha,\rho)$ for alpha between $1$ and
$10$ (left panel), then between $10$ and $100$ (right panel)
and $\rho=0.8,1.0,1.2$. The estimate of $\alpha$ was
computed using the $GLD(\alpha,\rho,F_{\mathrm{dir}})$, then averaged 
over all samples. The relative bias was calculated as the difference 
between the mean estimate and the true value of $\alpha$, 
divided by the true value of $\alpha$. Results are plotted as red
points. The results for the opposite experiment
(i.e. simulating the $GLD(\alpha,\rho,F_{\mathrm{dir}})$, and
  estimating using the $LD(\alpha,\rho)$) are plotted as green points.}}
\label{fig:ba1100}
\end{figure}

\begin{figure}[!ht]
\begin{center}
\includegraphics[width=4in]{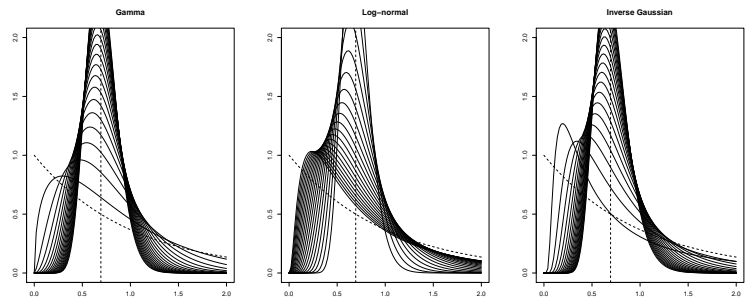}
\end{center}

\caption{
{\bf Densities of Gamma, Log-normal, and Inverse Gaussian.}  
All densities have been rescaled to have unit growth rates. The dashed
curve is the density of the exponential distribution with rate
$1$. The dashed vertical line locates the Dirac distribution at $\log
2$.
}
\label{fig:GALNIG}
\end{figure}

\begin{figure}[!ht]
\begin{center}
\includegraphics[width=4in]{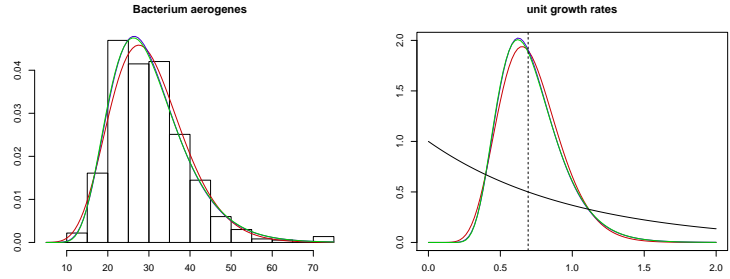}
\end{center}

\caption{
{\bf Adjusted distributions for
Kelly and Rahn's data on Bacterium aerogenes 
\cite[Table~2, p.~149]{KellyRahn32}.}
On the left panel, the histogram of the data, and the three densities
are superposed; the Gamma distribution appears in red, the Log-normal
distribution in blue, the Inverse Gaussian in green. The blue and
green curves are very close. On the right panel, the densities have
been rescaled to unit growth rate. The dashed curve is the density of
the exponential distribution, the dashed vertical line locates the
Dirac distribution at $\log 2$.
}
\label{fig:BA}
\end{figure}

\begin{figure}[!ht]
\begin{center}
\includegraphics[width=3.5in]{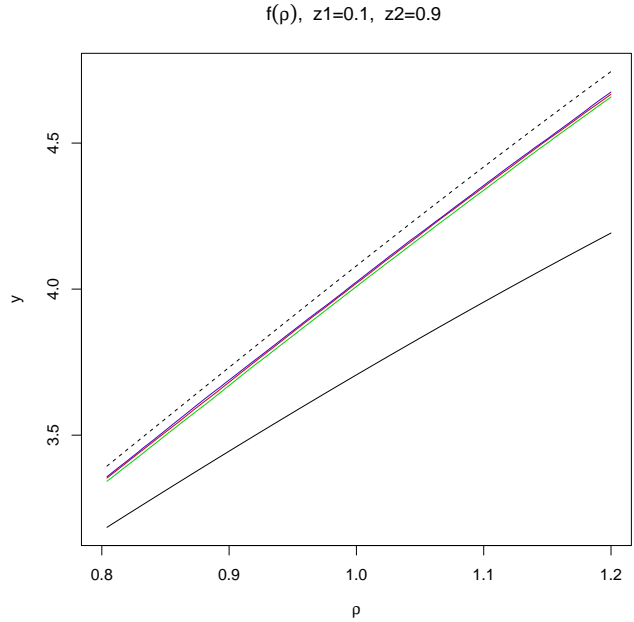}
\end{center}

\caption{
{\bf Ratios for GF estimators of the relative fitness $\rho$.}
Ratios $f_{z_1,z_2}(\rho)=\frac{h_\rho(z_1)-1}{h_\rho(z_2)-1}$ as
functions of $\rho$, for
$z_1=0.1$ and $z_2=0.9$. The ratios
depend on the division time distribution: exponential (solid
black), Dirac (dashed black), Gamma (red), Log-normal (blue), Inverse
Gaussian (green). The realistic distributions are close together, and 
closer to the Dirac case than to the exponential case. This explains
why the classical Luria-Delbr\"{u}ck model induces a positive bias on the
estimation of $\rho$, and why the Dirac model yields better results.
}
\label{fig:hr}
\end{figure}
\clearpage
\section*{Tables}
\begin{table}[!ht]
\caption{
\bf{Mean biases on estimates of alpha and rho.}}
\begin{tabular}{|c|c|c|c|c|}
\hline
parameters&$\hat{\alpha}_{\mathrm{dir}}$&$\hat{\alpha}_{\mathrm{exp}}$&
$\hat{\rho}_{\mathrm{dir}}$&$\hat{\rho}_{\mathrm{exp}}$\\\hline
$\alpha=1\,,\;\rho=0.8$&$0.011$&$0.016$&$0.003$&$0.083$\\
$\alpha=1\,,\;\rho=1.0$&$0.008$&$0.011$&$0.002$&$0.163$\\
$\alpha=1\,,\;\rho=1.2$&$0.010$&$0.010$&$0.003$&$0.278$\\\hline
$\alpha=4\,,\;\rho=0.8$&$0.061$&$0.064$&$0.000$&$0.073$\\
$\alpha=4\,,\;\rho=1.0$&$0.061$&$0.041$&$-0.001$&$0.137$\\
$\alpha=4\,,\;\rho=1.2$&$0.067$&$0.033$&$-0.001$&$0.235$\\\hline
$\alpha=8\,,\;\rho=0.8$&$0.166$&$0.028$&$0.002$&$0.056$\\
$\alpha=8\,,\;\rho=1.0$&$0.142$&$-0.059$&$-0.001$&$0.107$\\
$\alpha=8\,,\;\rho=1.2$&$0.156$&$-0.079$&$0.000$&$0.188$\\\hline
\end{tabular}
\begin{flushleft}
For each of the 9 sets of parameters (left column), 10000 samples of
size 100 of the $GLD(\alpha,\rho,F)$ were simulated, $F$ being the
Log-normal distribution adjusted on Kelly and Rahn's data. The
estimates of $\alpha$ and $\rho$ were calculated with the two models
Dirac and exponential. The estimated bias is the 
mean difference between the estimate and the true value. Biases on
$\rho$ with the classical model (rightmost column) are of order 10\% to 20\%.
\end{flushleft}
\label{tab:bias}
\end{table}

\begin{table}[!ht]
\caption{
\bf{Proportion of success for 95\% confidence intervals.}}
\begin{tabular}{|c|c|c|c|c|}
\hline
parameters&$\hat{\alpha}_{\mathrm{dir}}$&$\hat{\alpha}_{\mathrm{exp}}$&
$\hat{\rho}_{\mathrm{dir}}$&$\hat{\rho}_{\mathrm{exp}}$\\\hline
$\alpha=1\,,\;\rho=0.8$&$0.950$&$0.950$&$0.951$&$0.960$\\
$\alpha=1\,,\;\rho=1.0$&$0.951$&$0.952$&$0.946$&$0.954$\\
$\alpha=1\,,\;\rho=1.2$&$0.954$&$0.953$&$0.950$&$0.957$\\\hline
$\alpha=4\,,\;\rho=0.8$&$0.945$&$0.947$&$0.949$&$0.896$\\
$\alpha=4\,,\;\rho=1.0$&$0.947$&$0.950$&$0.947$&$0.845$\\
$\alpha=4\,,\;\rho=1.2$&$0.949$&$0.952$&$0.946$&$0.787$\\\hline
$\alpha=8\,,\;\rho=0.8$&$0.948$&$0.956$&$0.950$&$0.878$\\
$\alpha=8\,,\;\rho=1.0$&$0.948$&$0.953$&$0.952$&$0.805$\\
$\alpha=8\,,\;\rho=1.2$&$0.944$&$0.948$&$0.949$&$0.695$\\\hline
\end{tabular}
\begin{flushleft}
For each of the 9 sets of parameters (left column), 10000 samples of
size 100 of the $GLD(\alpha,\rho,F)$ were simulated, $F$ being the
Log-normal distribution adjusted on Kelly and Rahn's data. The 95\%
confidence intervals for $\alpha$ and $\rho$ were calculated with the two models
Dirac and exponential. The entries of the table are
proportions of the 10000 samples for which
the true value is in the confidence
interval. A result close to $0.95$ indicates a satisfactory
estimation.
\end{flushleft}
\label{tab:success}
\end{table}

\begin{table}[!ht]
\caption{
\bf{Confidence intervals for published data sets.}}
\begin{tabular}{|l|c|c|c|c|c|}
\hline
reference&size&
$\hat{\alpha}_{\mathrm{dir}}$&$\hat{\alpha}_{\mathrm{exp}}$&
$\hat{\rho}_{\mathrm{dir}}$&$\hat{\rho}_{\mathrm{exp}}$\\\hline
Luria \& Delbr\"uck A\cite{LuriaDelbruck43}&42&
$[5.30;9.09]$&$[5.22;8.89]$&$[0.77;1.19]$&$[0.82;1.34]$\\\hline
Luria \& Delbr\"uck B\cite{LuriaDelbruck43}&32&
$[0.34 ;1.03 ]$&$[0.35 ;1.04 ]$&$[0.21 ;0.76 ]$&$[0.18 ;0.80 ]$\\\hline
Boe et al. \cite{Boeetal94}&1102&
$[0.64 ;0.77 ]$&$[0.65 ;0.77 ]$&$[0.69 ;0.83 ]$&$[0.73 ;0.91 ]$\\\hline
Roshe \& Foster\cite{RoscheFoster00}& 52&
$[1.03 ;1.98 ]$&$[1.03 ;1.98 ]$&$[1.02 ;4.25 ]$&$[0 ;12.12 ]$\\\hline
Zheng \cite{Zheng02}& 30&
$[6.79 ;13.21 ]$&$[6.66 ;12.78 ]$&$[0.64 ;1.02 ]$&$[0.67 ;1.11 ]$\\\hline
\end{tabular}
\begin{flushleft}
For 5 published data sets, the 95\% confidence intervals on $\alpha$
and $\rho$ were calculated with the two models
Dirac and exponential.
\end{flushleft}
\label{tab:published}
\end{table}

\begin{table}[!ht]
\caption{
{\bf{Kolmogorov-Smirnov goodness-of-fit tests for
    published data sets.}}}
\begin{tabular}{|l|c|c|c|c|}
\hline
&\multicolumn{2}{|c|}{Dirac model}
&\multicolumn{2}{|c|}{Exponential model}
\\\hline
reference&distance&p-value&distance&p-value\\\hline
Luria \& Delbr\"uck A\cite{LuriaDelbruck43}&0.055&1.000&0.057&0.999\\\hline
Luria \& Delbr\"uck B\cite{LuriaDelbruck43}&0.069&0.998&0.055&1.000\\\hline
Boe et al. \cite{Boeetal94}&0.015&0.955&0.006&1.000\\\hline
Roshe \& Foster\cite{RoscheFoster00}&0.046&1.000&0.049&1.000\\\hline
Zheng \cite{Zheng02}& 0.063&1.000&0.070&0.997\\\hline
\end{tabular}
\begin{flushleft}
The Kolmogorov-Smirnov distance between the
sample and the adjusted distribution was calculated for the two models
Dirac and exponential. The parameters of the adjusted models were
estimated from the data by the GF method. Since the adjusted model
used estimations from the data, the p-value can only be taken as an 
indication. Calculations were made using the
 R package \verb+dgof+ \cite{ArnoldEmerson11}. 
\end{flushleft}
\label{tab:publishedKS}
\end{table}

\begin{table}[!ht]
\caption{
\bf{Characteristics of three families of distribution}}
\begin{tabular}{|l|c|c|c|}
\hline
Distribution&Gamma& Log-normal& Inverse Gaussian\\\hline
parameters& $GA(a,\lambda)$&$LN(\mu,\sigma)$&$IG(\mu,\lambda)$\\\hline
PDF&
$\displaystyle{\frac{t^{a-1}\lambda^a}{\Gamma(a)}\mathrm{e}^{-\lambda
    t}}$&
$\displaystyle{\frac{1}{x\sqrt{2\pi\sigma^2}}\mathrm{e}^{-\frac{\log(t) -
    \mu)^2}{2\sigma^2}}}$&
$\displaystyle{\left(\frac{\lambda}{2\pi t^3}\right)^{1/2}
\mathrm{e}^{-\frac{\lambda(t-\mu)^2}{2\mu^2 t}}}$\\\hline
Laplace transform&
$\displaystyle{\left(\frac{\lambda}{s+\lambda}\right)^a}$&
numeric&
$\displaystyle{\exp\left(\frac{\lambda}{\mu}\left(1-\sqrt{1+\frac{2\mu^2
        s}{\lambda}}\right)\right)}$\\\hline
growth rate&$\lambda(2^{1/a}-1)$&g.r. numeric&
$\displaystyle{\frac{\log 2}{\mu}+\frac{\log^2 2}{2\lambda}}$\\\hline
unit growth rate&
$GA(a,1/(2^{1/a}-1))$&
$LN(\mu+\log(\mathrm{g.r.}),\sigma)$&
$\displaystyle{IG(\log 2+\frac{\mu\log^2 2}{2\lambda},
\frac{\lambda\log 2}{\mu}+\frac{\log^2 2}{2})}$\\\hline
\end{tabular}
\begin{flushleft}
For Gamma, Log-normal and Inverse Gaussian distributions, 
the notation of parameters, the probability distribution function
(PDF), the Laplace transform, the growth rate, and the scaling 
for unit growth rate are given.
\end{flushleft}
\label{tab:GALNIG}
\end{table}

\begin{table}[!ht]
\caption{
\bf{Adjusted distributions for
Kelly and Rahn's data on Bacterium aerogenes 
\cite[Table~2, p.~149]{KellyRahn32}.}}
\begin{tabular}{|l|c|c|c|}
\hline
Distribution&Gamma& Log-normal& Inverse Gaussian\\\hline
parameters&$a=11.18\,,\;\lambda=15.64$&
$\mu=-0.38\,,\;\sigma=0.30$&
$\mu=0.72\,,\;\lambda=7.50$\\\hline
Kolmogorov-Smirnov&$0.693$&$0.919$&$0.874$\\\hline
Anderson-Darling&$0.505$&$0.852$&$0.813$\\\hline 
\end{tabular}
\begin{flushleft}
A maximum likelihood estimation of the parameters
on the data led to one particular distribution in each family, 
that was rescaled to unit growth rate. The parameters of the rescaled
distribution are given, together with the p-values for the
Kolmogorov-Smirnov and Anderson-Darling goodness-of-fit tests.
\end{flushleft}
\label{tab:BA}
\end{table}

\end{document}